# ON THE GLOBAL STABILITY OF MAGNETIZED

# ACCRETION DISKS.

## II. VERTICAL AND AZIMUTHAL MAGNETIC FIELDS

CHARLES CURRY and RALPH E. PUDRITZ

Department of Physics and Astronomy

McMaster University, Hamilton, Ontario L8S 4M1, Canada

*Email:curry@jabba.physics.mcmaster.ca,pudritz@physics.mcmaster.ca*

## ABSTRACT

We investigate the global stability of a differentially rotating fluid shell threaded by vertical and azimuthal magnetic fields to linear, axisymmetric perturbations. This system, which models a thick accretion disk in the vicinity of its midplane, is susceptible to the Velikhov-Chandrasekhar (VC) instability in the absence of the azimuthal field. In most cases, the azimuthal field tends to stabilize the VC instability, although strong fields (Alfvén speed of order the characteristic rotational speed in our incompressible model) are required for complete stabilization. Stability diagrams are constructed, indicating critical values of the two fields for instability. We find an additional strong field instability that arises when the azimuthal Alfvén speed exceeds the characteristic rotational speed. This instability, in the case of a freely bounded configuration, has certain similarities to the sausage instability for interpenetrating fields in plasma physics, and may be important for very massive disks or filamentary molecular clouds. An application to the L1641 region in Orion A is briefly discussed. Finally, we find that the effect of a radially varying vertical field (without an azimuthal field) is mainly stabilizing.

*Subject headings:* accretion, accretion disks - instabilities - ISM: magnetic fields - MHD

## 1. INTRODUCTION

One of the more interesting recent developments in the theory of accretion disks was the discovery of virulent instabilities that develop only in the presence of magnetic fields. Balbus & Hawley (1991) (hereafter BH) showed that a Keplerian disk in a state of pure rotation threaded by a weak axial field was subject to a local instability whose growth rate was on the order of the local rate of rotation. In a previous paper, we examined the global counterpart of this instability, the Velikhov-Chandrasekhar (VC) instability, showing that growth persisted at comparable rates (Curry, Pudritz, & Sutherland 1994, hereafter CPS). There remain serious questions, however, concerning how the instability is affected as models are augmented by additional physics. In particular, the influence of the more complicated magnetic field structures expected to exist in protostellar, CV, and AGN disks has yet to be carefully addressed. In this paper, we extend the model of CPS to include disks with radially varying vertical ($B_z$) and azimuthal ($B_\phi$) fields.

There are many reasons to expect an azimuthal field to be an important, sometimes dominant, magnetic field component in accretion disks. Strong differential rotation can generate $B_\phi$ from a radial field component $B_r$, which can itself be created either by dynamo processes or accretion. The BH instability has been shown to generate strong $B_r$ and $B_\phi$ from an initially weak $B_z$ (Hawley, Gammie, & Balbus 1995). If $B_z$ is inherited from the central object or the interstellar medium, disk torques can convert it directly to $B_\phi$. Thus it is most likely that all three components of **B** are dynamically important for most types of disks. Of course, as has been made clear by all work following BH, one needs very little initial $B_z$ in order for that component to be "dynamically important."

The observational evidence for $B_\phi$ in protostellar disks is at the present time quite sparse, mainly due to uncertainties about the nature of the detected disks themselves. Recent mid-infrared spectropolarimetry of high-mass star-forming regions by Aitken et al. (1993) revealed a high correlation between objects with elongated molecular disk-like structures (numbering 10 in their sample) and magnetic fields oriented along the long axis of the disk (7 of these 10). The authors claim this as evidence for a predominantly azimuthal field structure in these regions. One should note, however, that the objects in question are $10^3$ to $10^4$ AU in extent, with masses $\sim 10^3 M_\odot$, and so are not likely to represent Keplerian accretion disks. As evidence for large-scale rotation is lacking, they may in fact be self-gravitating toroids or "pseudodisks," supported to some extent by the field itself (Galli & Shu 1993).

Keplerian disks with magnetic field components in both the azimuthal and vertical directions have been actively studied as possible sources of centrifugally driven winds and outflows (e.g. Blandford & Payne 1982, Uchida & Shibata 1985, Pelletier & Pudritz 1992). This suggests an additional motivation for the present study: to determine whether the various equilibria assumed in models of magnetically driven outflows are stable. A first step in this direction was taken recently by Lubow, Papaloizou & Pringle

(1994).

In a different context, Galactic center molecular disk observations (Genzel 1989, Hilderbrand et al. 1990) indicate that $B_r \sim B_\phi \sim 1$ mG, with a somewhat weaker $B_z$. This is in contrast to the larger-scale field structure (i.e. the inner 70 pc of the Galaxy), which is almost purely vertical, i.e. perpendicular to the Galactic plane. Wardle & König (1990) have modeled this region using a self-similar magnetized disk model, under the assumption that the inner field structure results from advection of the large-scale field by inflowing matter, with differential rotation subsequently leading to a strong azimuthal component. As observations of other galactic nuclei and AGN are still not able to resolve the inner disks, much less any associated magnetic field structure, it would be unwise to speculate further along these lines. However, since the inner regions of AGN are expected to possess "thick" rather than thin Keplerian disks, we use the same equilibrium sequence as introduced in CPS; namely, one in which radial pressure gradients oppose the central gravity for non-Keplerian rotation laws. The situation examined in the present paper is even more interesting, however, since radial *magnetic* gradients are also present.

As a final possible application of the work presented here, we cite evidence that elongated filaments of gas in molecular clouds are associated with helical velocity and magnetic fields (Bally 1989). The latter are indicative of the simultaneous presence of $B_\phi$ and $B_z$. The model employed in this paper, although formulated primarily for accretion disks, yields interesting results in the parameter range expected to hold in such regions. In particular, we find a new instability which sets in when the azimuthal Alfvén speed is greater than the rotational speed.

We defer to a later section a detailed description of previous work on the effect of $B_\phi$ on the VC and BH instabilities, but it is of use to review here what is known generally about the stability of rotating configurations with azimuthal field. Since we do not attempt to account for the vertical structure of the disk in this study, the following discussion is restricted to purely radial distributions of angular velocity and magnetic field. The central question is this: given a rotation profile $\Omega(r)$ and field distribution $B_\phi(r)$, $B_z(r)$, can one predict, even locally, whether a configuration is stable to infinitesimal perturbations in the fluid quantities ? What is needed is a necessary *and* sufficient criterion for stability, such as exists for purely vertical and purely azimuthal fields. These are, respectively;

$$\frac{d\Omega^2}{dr} \geq 0$$

and

$$\frac{1}{r^2}\frac{d}{dr}(r^2\Omega)^2 - \frac{r^2}{4\pi\rho}\frac{d}{dr}\left(\frac{B_\phi}{r}\right)^2 \geq 0. \tag{1.1}$$

The first criterion is due to Chandrasekhar (1960), and the second to Michael (1954). On the other hand, for the combined fields in the absence of rotation, the relevant stability criterion is (Chandrasekhar 1961)

$$\frac{d}{dr}(rB_\phi)^2 \leq 0.$$

In the most general case of combined fields with rotation, no similar criterion is known. *Sufficient* criteria are available, however; one of these is (Howard & Gupta 1962, Dubrulle & Knobloch 1993, Kumar, Coleman, & Kley 1994):

$$r\frac{d\Omega^2}{dr} - \frac{2B_\phi}{4\pi\rho r^2}\frac{d}{dr}(rB_\phi) \geq 0. \tag{1.2}$$

Sufficient criteria can only be regarded as incomplete guides to the global stability of systems; the inherent limitations of criterion (1.2) will be made manifest later on in the paper.

As to the actual distribution of $B_\phi$ and $B_z$ across the disk, there seem to be very few restrictions at this time[1]. By considering power-law distributions in these two field components, we hope to cover a range of plausibility.

It is important to emphasize that the goal of this series of papers is not to replace the many local analyses that exist in the literature. Rather, a model such as we utilize below, while idealized and unrealistic in many respects, highlights intrinsically global behavior which will not be discovered in any local analysis. Examples of this found in the present work and in CPS are effects which involve coherent motions over large portions of the disk, and phenomena modified or even enhanced by the presence of a disk boundary, imperfectly modelled though it may be. Thus the present work complements, not replaces, existing local analyses.

The format of the paper is as follows. The equilibrium state is described in §2, and the perturbations to this state in the following section. Quantitative results for the combined effect of azimuthal and constant vertical fields are presented in §4, and those for a radially varying vertical field in §5. In the final section, our results are compared with those of other investigators, and we make some additional comments on a new instability found in §4, before giving a final summary. Technical details of the calculations may be found in the four appendices.

## 2. THE EQUILIBRIUM

### 2.1. Basic Equations

The equilibrium was described in detail in CPS, and has also been employed in stability analyses of thick, pressure-supported disks; see, e.g., Blaes & Glatzel (1986),

---

[1]Because an equilibrium $B_r$ immediately implies a time-dependent, growing $B_\phi$ (BH) which destroys the time-invariance of the resulting equations, we ignore this field component in the present work.

Sekiya & Miyama (1988), and Jaroszyński (1988). The model is a simplified form of the "thick torus" model for AGN (see, e.g., Paczyński & Wiita 1980), supplemented by gradients of magnetic field pressure, but lacking vertical structure. It should therefore adequately describe a small region straddling the midplane of a real disk, *with radial gas and magnetic pressure support taken fully into account.* Thus the equilibrium is *not* that of a Keplerian disk, although this case is naturally included in the equilibrium sequence (CPS).

Consider a cylindrical shell of homogeneous, incompressible, ideal MHD fluid, of infinite extent in the $z$-direction, rotating about the $z$-axis in the Newtonian point-mass potential $\Psi = -GM/r$. The purely radial dependence of the potential is justified if, at every radius $r$, the vertical scale height of the "disk" $H \ll r$, so that there is little variation of $\Psi$ with $z$. The stationary solution of the MHD equations depends only on the radial coordinate, $r$. To calculate explicit quantities of interest, we take the following power-law dependences for the angular velocity, azimuthal, and vertical magnetic fields, respectively:

$$\Omega(r) = \Omega_0 \left(\frac{r}{r_0}\right)^{-a}, \quad B_\phi(r) = B_{\phi 0} \left(\frac{r}{r_0}\right)^{-b+1}, \quad B_z(r) = B_{z0} \left(\frac{r}{r_0}\right)^{-c+1}, \qquad (2.1)$$

where $\Omega_0, B_{\phi 0}, B_{z0}, a, b, c,$ and $r_0$ are constants. As in CPS, we consider the effect of both rigid and free boundaries. In the latter case, **B** is supposed to permeate the regions both to the interior and exterior of the shell, as well as within the fluid. Using equations (2.1), the radial component of the equation of motion becomes (Appendix A)

$$\frac{p'}{\rho} = r_0 \Omega_0^2 \left(\frac{r}{r_0}\right)^{1-2a} - \frac{GM}{r^2} - \frac{1}{r_0}\left[(2-b)V_{\phi 0}^2\left(\frac{r}{r_0}\right)^{1-2b} + (1-c)V_{z0}^2\left(\frac{r}{r_0}\right)^{1-2c}\right], \quad (2.2)$$

where the prime symbol $\equiv d/dr$, $V_{\phi 0, z0}^2 \equiv B_{\phi 0, z0}^2/4\pi\rho$ are the azimuthal ($\phi$) and vertical ($z$) Alfvèn speeds at $r_0$, and $M$ is the mass of the central object (self-gravity is ignored).

Inspection of equation (2.2) shows that the magnetic terms aid rotation and oppose the central gravity if $b > 2$ and/or $c > 1$, and vice-versa if $b < 2$ and/or $c < 1$. As in CPS, we consider configurations in which the gas pressure vanishes at the boundaries, and identify $r_0$ with the gas pressure maximum, where $p' = 0$. Equation (2.2) then gives

$$\frac{GM}{r_0^2} = r_0 \Omega_0^2 - [(2-b)V_{\phi 0}^2 + (1-c)V_{z0}^2]/r_0. \qquad (2.3)$$

This is merely a statement of radial magnetostatic equilibrium at the pressure maximum. In order for $r_0$ to be a maximum, we must have $p''(r_0) < 0$. From equation (2.2), this requires

$$(2b-3)(b-2)\overline{V}_{\phi 0}^2 + (2c-3)(c-1)\overline{V}_{z0}^2 + 2a - 3 > 0, \qquad (2.4)$$

where an overline indicates that the Alfvén speeds are now scaled with respect to $r_0\Omega_0$, the circular speed at $r_0$. Note that the above gives $a > 3/2$ when $V_{\phi 0} = 0$ and $c = 1$, as expected. Condition (2.4) should be satisfied for each equilibrium we examine.

Integrating equation (2.2) and eliminating $GM$ via equation (2.3), one obtains the stationary pressure distribution

$$\frac{p}{\rho} = \frac{\mu^2}{2} + \frac{1}{r} - 1 + \frac{1 - r^{-2(a-1)}}{2(a-1)} - (2-b)\overline{V}_{\phi 0}^2 \left[\frac{1}{r} - 1 + \frac{1 - r^{-2(b-1)}}{2(b-1)}\right]$$
$$- (1-c)\overline{V}_{z0}^2 \left[\frac{1}{r} - 1 + \frac{1 - r^{-2(c-1)}}{2(c-1)}\right], \qquad (2.5)$$

where we have chosen our units such that $r_0 = \Omega_0 = 1$, and where $\mu^2$ is a constant equal to the ratio of thermal to kinetic energy at $r_0$. We assume, as in CPS, that the gas and magnetic pressures remain finite as $r_2 \to \infty$; this implies that $a, b, c \geq 1$.

## 2.2. *Special Cases*

### 2.2.1. $B_z = constant$

Much of this paper is based on the particular case of a constant vertical field; i.e. $c = 1$. Then equation (2.3) and inequality (2.4) lead to the inequalities (in dimensionless units)

$$1 - (2-b)V_{\phi 0}^2 > 0,$$
$$(2b-3)(b-2)V_{\phi 0}^2 + 2a - 3 > 0,$$

where we have dropped the overlines on the Alfvén speeds for convenience. Considering all possible values of $a$ and $b$ leads to the conclusion that only certain values of $V_{\phi 0}$ are permitted for a given $(a, b)$.

In the case of rigid boundaries, the inner and outer boundaries of the fluid are determined by the zeros of equation (2.5). For free boundaries, this is still true provided that $\mathbf{B}$ is continuous across the boundaries, and we shall assume that this is the case. Thus a given model is fixed by choosing $r_2/r_1$, $a$, $b$, and $V_{\phi 0}$. CPS found a monotonic increasing dependence of the VC instability growth rate on $r_2/r_1$, with a maximum at $r_2/r_1 \gtrsim 100$; thus, we choose $r_2/r_1 = 100$ as a fiducial value for all calculations in this paper. The zeros of equation (2.5) can be positive, negative, or complex. The latter two (unacceptable) possibilities can occur even for $(a, b, V_{\phi 0})$ obeying the above inequalities. We therefore conducted a three-parameter search for acceptable equilibria; the results are summarized in Figures 1a and b.

In Fig. 1a, various critical values of the azimuthal Alfvén speed are denoted by $V_1$, $V_2$, $V_3$, ...; each is a function of $a$ and $b$. Although we calculated equilibria for all $a, b$ in the range $1 \leq (a, b) \leq 3$, Fig. 1b shows only $3/2 \leq (a, b) \leq 2$. We will restrict consideration for most of the paper to this range, since it reduces exactly to the equilibrium of CPS when $B_\phi = 0$. The upper surface in Fig. 1b represents $V_5(a > b)$ and $V_6(a < b)$. Note that for a given $a$, $3/2 < a < 2$, $V_5 > V_6$.

Fig. 1.— (a) Allowed regions and limiting azimuthal Alfvén speeds in the $(a, b)$ plane, obtained from solution of the equilibrium equation (2.5) where $p = 0$, with $r_2/r_1 = 100$. Equilibria for values of $a$ and $b$ lying in the shaded region and along dashed lines are not allowed for any $V_{\phi 0}$. (b) 3D plot of allowed equilibria. Only the range $3/2 \leq (a, \ b) \leq 2$ is shown. Permissible $V_{\phi 0}$ for a given $a, b$ lie between the upper and lower surfaces. The upper surface represents $V_5(a > b)$ and $V_6(a < b)$.

There is another interesting property of the equilibrium relation (2.5): when $b = 2$ and $c = 1$, we get the equilibrium of CPS. It may be checked that, for the power law fields we assume, this is the unique solution for which the current density, $\mathbf{J} = \nabla \times \mathbf{B}/\mu_0$, vanishes. Hence, the value of $a$ is restricted to $3/2 \leq a \leq 2$, just as in that study, and the location of the inner radius given by equation (2.9) of CPS. Since this special case allows us to examine the effect of the azimuthal field without the added complication of a current, we will assume $b = 2$, $c = 1$ (corresponding to $B_\phi \sim r^{-1}$, $B_z = $ constant) when considering free boundaries in the sections to follow.

### 2.2.2.   $B_z = B_z(r)$, $B_\phi = 0$

In this case equations (2.3) and (2.4) give

$$1 - (1 - c)V_{z0}^2 > 0,$$
$$(2c - 3)(c - 1)V_{z0}^2 + 2a - 3 > 0.$$

As above, these inequalities and equation (2.5) impose restrictions on allowed equilibria; these are summarized in Figures 2a and 2b. We now examine perturbations to the above-described equilibria.

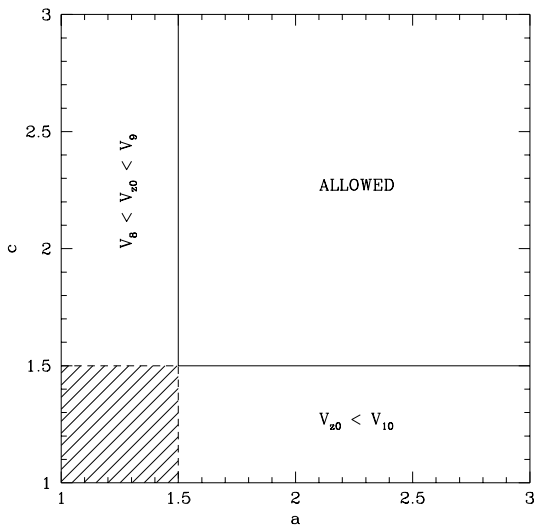

Fig. 2.— (a) Allowed regions and limiting vertical Alfvén speeds in the $(a, c)$ plane. (b) 3D plot of allowed equilibria. For the range of $a$ shown, restrictions on $V_{z0}$ apply only for $1 \leq c \leq 3/2$. The upper surface represents $V_{10}$. See text and Fig. 1 caption for details.

## 3. THE PERTURBATIONS

### 3.1. *The Perturbation Equations*

We now consider the response of the above equilibrium state to small, axisymmetric, Eulerian perturbations of the form

$$\delta X(r, z, t) = \delta X(r)e^{i(kz + \omega t)}, \tag{3.1}$$

where $X$ is any physical variable, and $k$ and $\omega$ are the vertical wavenumber and frequency of the perturbation, respectively. Substituting the forms $X + \delta X$ along with equation (2.1) into the ideal MHD equations, linearizing, and eliminating all variables in favor of the radial velocity perturbation (see Appendix A for details), one obtains a second-order differential equation in $\delta u_r$:

$$\frac{1}{r}[r\tilde{\omega}^2(\delta u_r)']' + q(r)\delta u_r = 0, \tag{3.2}$$

where

$$q(r) \equiv k^2 r \left[ \left( \Omega^2 - \frac{V_\phi^2}{r^2} \right)' - \frac{(V_z^2)'}{r^2} \right] \quad + \quad \frac{4k^2}{\tilde{\omega}^2} \left( \frac{kV_\phi V_z}{r} - \omega\Omega \right)^2$$
$$- \quad \tilde{\omega}^2 \left( k^2 + \frac{1}{r^2} \right),$$

$$\tilde{\omega}^2 \equiv \omega^2 - k^2 V_z^2(r),$$

and $V_{\phi,z}(r) = B_{\phi,z}(r)/\sqrt{4\pi\rho}$ are the azimuthal and vertical Alfvén speeds. The power-law form of the above (in dimensionless units) is

$$
\begin{aligned}
q(r) =\ & 2k^2 \left[ b V_{\phi 0}^2 r^{-2b} - a r^{-2a} + (c-1) V_{z0}^2 r^{-2c} \right] \\
& + \frac{4k^2}{\tilde{\omega}^2} (k V_{\phi 0} V_{z0} r^{1-b-c} - \omega r^{-a})^2 - \tilde{\omega}^2 \left( k^2 + \frac{1}{r^2} \right).
\end{aligned}
\tag{3.3}
$$

An alternative form of the perturbation equation useful for analytic purposes is obtained via the transformation

$$\psi \equiv (r\tilde{\omega}^2)^{1/2} \delta u_r,$$

whence equation (3.2) becomes

$$\psi'' = k^2 Q(r) \psi, \tag{3.4}$$

with

$$
\begin{aligned}
Q(r) =\ & \frac{2}{\tilde{\omega}^2} \left\{ a r^{-2a} - b V_{\phi 0}^2 r^{-2b} - (c-1) V_{z0}^2 r^{-2c} - \frac{2}{\tilde{\omega}^2} (k V_{\phi 0} V_{z0} r^{1-b-c} - \omega r^{-a})^2 \right\} \\
& + \left( 1 + \frac{1}{k^2 r^2} \right) - \frac{1}{2k^2} \left[ \frac{1}{2} \frac{(r\tilde{\omega}^2)'^2}{(r\tilde{\omega}^2)^2} - \frac{(r\tilde{\omega}^2)''}{r\tilde{\omega}^2} \right].
\end{aligned}
\tag{3.5}
$$

When discussing free-boundary configurations, one must consider the form of the vacuum field perturbations in addition to those within the fluid. We will restrict ourselves to the *current-free* case, i.e. $b = 2, c = 1$, since then the perturbed magnetic field in the interior ($r < r_1$, denoted by subscript $i$) and exterior ($r > r_2$, subscript $o$) regions is completely specified by a scalar potential $\chi$, such that

$$\delta \mathbf{B}_{i,o} = B_z \nabla \chi_{i,o},$$

$$\chi_i(\varpi) = c_1 I_0(\varpi), \quad \chi_o(\varpi) = c_2 K_0(\varpi), \tag{3.6}$$

where $\chi_{i,o} = \chi_{i,o}(r) e^{i(kz+\omega t)}$, $\varpi \equiv |k|r$, and $I_0$ and $K_0$ are modified Bessel functions of order zero.

### 3.2. *The Boundary Conditions*

We solve equation (3.2) subject to both rigid and free boundary conditions (BCs). The former are

$$\delta u_r(r_1) = \delta u_r(r_2) = 0.$$

Free BCs require the continuity of Lagrangian perturbations of the total normal stresses and magnetic flux across the boundaries. In the cylindrical geometry we are considering, both $B_\phi$ and $B_z$ are everywhere perpendicular to the surface normal **n**; thus **B** $\cdot$ **n** $= 0$, and provided that both fields are continuous across the boundaries, the appropriate BC is unchanged from the constant $B_z$ case; that is (CPS),

$$(\delta u_r)' + \left[ \frac{1}{r} + \frac{k^2}{\tilde{\omega}^2} \left( g_{eff} + k^2 V_{z0}^2 \frac{\chi_{i,o}}{\chi'_{i,o}} \right) \right] \delta u_r = 0.$$

The subscript $i$ applies at $r_1$, subscript $o$ at $r_2$.

For general power laws in $B_\phi(r), B_z(r)$, and in dimensionless units, the effective gravity is given by

$$g_{eff} = r^{1-2a} - \frac{1}{r^2} - (2-b)V_{\phi 0}^2 r^{1-2b} - (1-c)V_{z0}^2 r^{1-2c}.$$

For vanishing current, this becomes identical to the $g_{eff}$ of CPS; i.e. $g_{eff} = r^{1-2a} - 1/r^2$.

From equation (3.6) one finds

$$\left. \frac{\chi_i}{\chi'_i} \right|_{r=r_1} = \frac{1}{|k|} \frac{I_0(\varpi_1)}{I_1(\varpi_1)} \quad \text{and} \quad \left. \frac{\chi_o}{\chi'_o} \right|_{r=r_2} = -\frac{1}{|k|} \frac{K_0(\varpi_2)}{K_1(\varpi_2)},$$

where $\varpi_{1,2} = |k|r_{1,2}$.

## 4.   RESULTS: CONSTANT VERTICAL FIELD

The majority of our results have been obtained for the special case $V_z = $ constant.

### 4.1.   *The Case of $a = b$*

When $a = b$, the rotation frequency $\Omega$ and its magnetic analog, $V_\phi/r$, have an identical scaling with radius. Equation (3.2) with (3.3) becomes

$$\frac{1}{r}[r(\delta u_r)']' + Q(r)\delta u_r = 0, \tag{4.1}$$

where

$$Q(r) = k^2(Er^{-2a} - 1) - \frac{1}{r^2}, \tag{4.2}$$

$$E \equiv \frac{2}{\tilde{\omega}^4}\left[a\tilde{\omega}^2(V_\phi^2 - 1) + 2(\Omega_A V_\phi - \omega)^2\right], \tag{4.3}$$

$\Omega_A \equiv kV_z$ is the Alfvén frequency, and where we have dropped the zero subscripts on $V_\phi$ and $V_z$ for convenience. The reader should note that these are *constants* throughout this section.

Equation (4.1) with equation (4.2) is identical in form to the perturbation equation examined in CPS; for rigid BCs, the two problems are formally identical. The eigenvalue $E$ is a known function of $a$ and $k$ (see CPS and ff. equation (4.8)), but here its definition in terms of $\omega$ differs. The latter are solutions of the quartic polynomial obtained from equation (4.3):

$$E\omega^4 - 2[E\Omega_A^2 + a(V_\phi^2 - 1) + 2]\omega^2 + 8\Omega_A V_\phi \omega + \Omega_A^2[E\Omega_A^2 + 2a(V_\phi^2 - 1) - 4V_\phi^2] = 0. \quad (4.4)$$

The eigenvalue spectrum for $E$ is infinite and every member is real and positive. For free BCs, the situation is complicated by the fact that $\omega$ appears in the BC itself. The problem is then no longer a standard Sturm-Liouville one and the introduction of $E$ is not a particularly useful calculational device. In fact, when $V_\phi \neq 0$, the resulting $E$ is always *complex*. This leads to some interesting consequences which will be discussed presently.[2] For both sets of BCs, the resulting $\omega$ occur in complex conjugate pairs (Frieman & Rotenberg 1960).

For simplicity, we begin by considering rigid BCs only. There are two special cases in which the roots of equation (4.4) have simple analytic forms. One is when $V_\phi = 1$, which will be examined in the next section. The other is when $a = 2$. In that case, two of the roots are always stable, and the other two are

$$\omega = \frac{V_\phi \pm (V_\phi^2 + E\Omega_A^2 - 2E^{1/2}\Omega_A)^{1/2}}{E^{1/2}}. \quad (4.5)$$

The critical Alfvén frequency for stability, where the imaginary part of equation (4.5) vanishes, is then

$$\Omega_{A,crit} = \frac{1 \pm (1 - V_\phi^2)^{1/2}}{E^{1/2}}. \quad (4.6)$$

Two points are worth noting. First, $V_\phi$ has a *stabilizing* influence here. This could not be predicted from the sufficient criterion (1.2) given in the introduction, since $b = 2 \Rightarrow (rB_\phi)' = 0$. Second, $\Omega_{A,crit}$ is an explicit function of $V_\phi$, and has *two* distinct nonzero solutions. That is, *the stability criterion is altered in the presence of an azimuthal field*, contrary to the claims of some recent investigators (see ff. §6.1). In the local limit, i.e. $k \to \infty$, $V_z \to 0$, we have $E \to r_1^{2a} = r_1^4$ (CPS); then equation (4.5) gives

$$\omega = \frac{1}{r_1^2}[V_\phi - (V_\phi^2 - 2\Omega_A r_1^2 + \Omega_A^2 r_1^4)^{1/2}]$$

for the growing unstable mode.

We solved equation (4.1) numerically, subject to both rigid and free boundary conditions, for a variety of $(a, b, V_\phi)$ allowed by the equilibrium. As in CPS, we use the WKB approximation when $V_z \lesssim 0.3$, since then the eigenfunctions $\delta u_r$ are so sharply peaked that numerical solutions are difficult to obtain.

---

[2]When $V_\phi = 0$ and for free BCs, as in CPS, it can be shown that the problem is still of Sturm-Liouville type, since $\omega^2$ is real and certain required conditions on the BC coefficients are satisfied; cf., Birkhoff & Rota (1989).

The principal results are as follows:

(i) The VC instability persists for all $V_\phi < 1$, $3/2 \leq a \leq 2$, but with reduced growth rate. This conflicts with the naive prediction based on the sufficient criterion (1.2), since here $(rB_\phi)' \geq 0$. The growth rate approaches zero as $V_\phi \to 1$.

(ii) The presence of the azimuthal field also changes the stability criterion itself. Growth is damped at both short and long wavelengths.

(iii) When $V_\phi > 1$, $a \neq 2$, a new instability sets in, increasing in growth rate as $V_\phi$. This large-field instability can be stabilized if $V_z$ is made sufficiently large.

(iv) All of the unstable modes propagate; i.e. the real part of $\omega$ is $\omega_R \sim kV_\phi V_z$.

(v) The mode structure is unchanged from CPS; i.e. there exists a finite, ordered spectrum of unstable modes, whose growth rates are inversely proportional to a positive power of $E$. For the remainder of this paper, we will restrict consideration to the fastest-growing, or $n = 0$, mode.

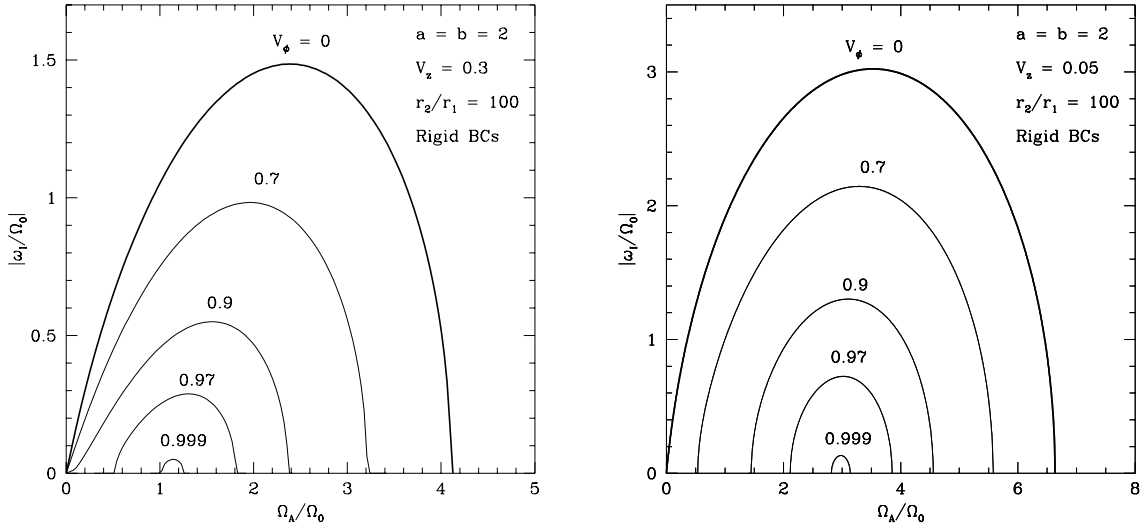

Fig. 3.— WKB growth rates as a function of Alfvén frequency $\Omega_A = kV_z$ for the fastest-growing ($n = 0$) mode, a range of $V_\phi$, and two different vertical field values: (a) $V_z = 0.3$; (b) $V_z = 0.05$.

In Figures 3a and b, we plot the dimensionless growth rate as a function of the Alfvén frequency $\Omega_A = kV_z$. These curves show directly the effect of azimuthal field on the VC instability. We have chosen $a = b = 2$, but the curves are similar for other $a = b$. To display the effect for both strong and weak axial fields, Fig. 3a has $V_z = 0.3$ and Fig. 3b, $V_z = 0.05$. Feature (i) is apparent in both figures; growth is clearly halted as $V_\phi \to 1$. In the presence of $B_\phi$, growth rates are reduced due to vertical motions induced by magnetic pressure gradients (Blaes & Balbus 1994) (in the absence of $B_\phi$, $\delta u_z \sim \delta p$; compare equation (3.1) of CPS and equation (A.4), Appendix A). The instability couples to (stable) inertial modes, reducing its efficacy. The additional stabilization provided by $V_\phi$ at shorter wavelengths (large $\Omega_A$) is also apparent in both figures. The physical explanation for this is the same as in CPS; namely, that the restoring stress on a fluid element is more effective for distortions of larger curvature, i.e. at short wavelengths (also, see below).

A new effect, the long-wavelength stabilization, is much more prominent in the weak axial field case (Fig. 3b). Even at $V_\phi \approx 0.7$, one sees stabilization at long wavelengths (small $\Omega_A$) for $V_z = 0.05$. This behavior is entirely due to the presence of toroidal field lines, which provide an additional return force on a fluid element at long wavelengths. This can be seen by an explicit calculation of the perturbed magnetic tension, i.e.

$$\frac{\delta(\mathbf{B} \cdot \nabla)\mathbf{B}}{4\pi} = \frac{1}{4\pi}\left(ikB_z\delta\mathbf{B} - \frac{2B_\phi\delta B_\phi}{r}\hat{\mathbf{r}}\right), \tag{4.7}$$

where we have assumed without loss of generality that $B_\phi \sim 1/r$. As $k \to 0$ in the $B_\phi = 0$ case, the tension vanishes, indicating that instability persists up to the longest wavelengths. However, the second RHS term is independent of $k$, so that for nonzero $B_\phi$ there exists an additional radial tension, which is always stabilizing. In addition, the effect is enhanced at small $B_z$. It is this behavior that we observe at small $\Omega_A$ in Fig. 3.

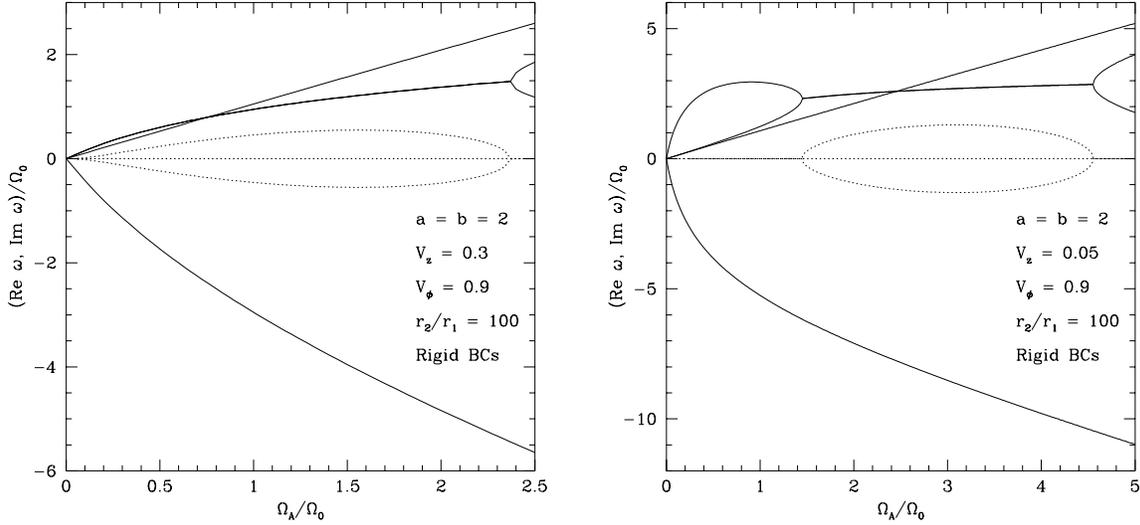

Fig. 4.— Real (solid lines) and imaginary (dotted) parts of the eigenfrequency $\omega$ as a function of Alfvén frequency for $V_\phi = 0.9$ and (a) $V_z = 0.3$; (b) $V_z = 0.05$. Other parameters are the same as in Fig. 3.

The real parts of all four roots for $\omega$ are shown in Figures 4a and b, for the same two values of $V_z$ and $V_\phi = 0.9$. The corresponding imaginary parts are shown as dotted lines. The unstable modes (one growing, one damping) are created out of two real modes which merge for intermediate values of $\Omega_A$.

Increasing $V_\phi$ to values in excess of 1 with $a = b = 2$ leads to no further instability. However, a new instability *does* occur for other values of $a$. The Keplerian case, for example, is shown in Figure 5. Each curve is labeled by its corresponding field values $V_\phi$, $V_z$. As $V_z$ is held fixed at 0.3 and $V_\phi$ increased, the peak growth rate increases (solid curves). Were it not for the equilibrium constraint $V_\phi \lesssim 1.42$ (see Fig. 1a), this growth would continue without bound as $V_\phi$ is increased. Now keeping $V_\phi$ fixed and increasing $V_z$ from 0.3 (dashed curves) leads to stabilization, until complete stability is achieved at $V_z \simeq 0.81$, implying $(V_z/V_\phi)_{crit} \simeq 0.57$. We consider this large-field instability further in §4.3.

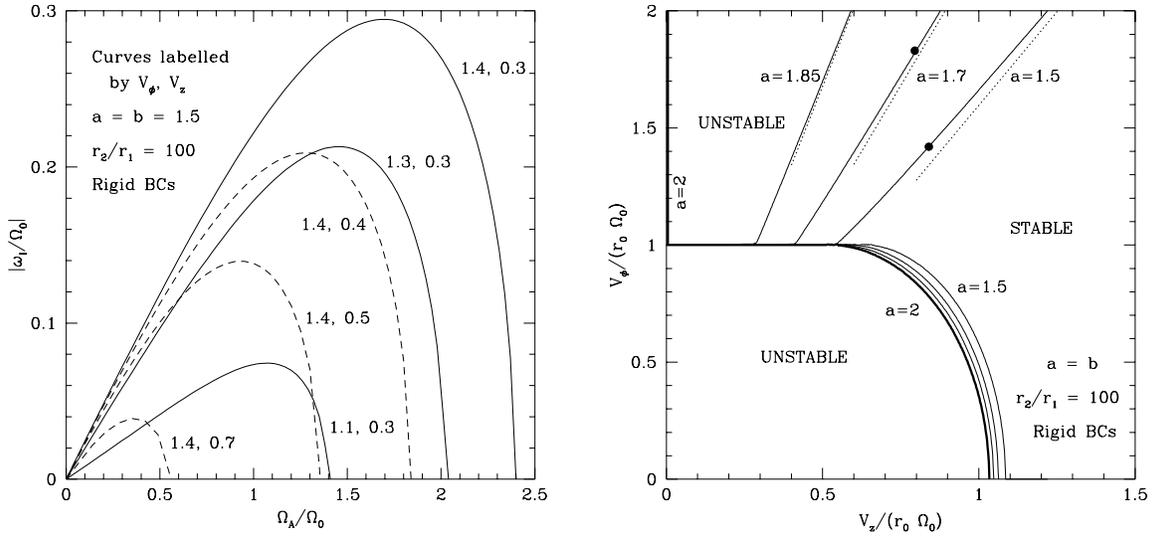

Fig. 5.— Growth rates of the large-field instability as a function of Alfvén frequency for Keplerian rotation and rigid BCs. Each curve is labelled by its corresponding $V_\phi$, $V_z$. Solid curves have $V_z = 0.3$, while dashed curves have $V_\phi = 1.4$. The chosen Alfvén speeds are consistent with the equilibrium constraint $V_\phi \leq 1.42$.

Fig. 6.— Critical stability curves, Im $\omega = 0$, in the $(V_\phi, V_z)$ plane for rigid BCs and (from right to left) $a = b = 1.5$, 1.7, 1.85, and 2. Growth rates increase from zero on both sides of each critical curve. The region at lower left is VC unstable for all $a$; the similar region at top left is large-field unstable for all $a$ except $a = 2$. The dotted lines are the slopes of the LFI found analytically from equation (4.11). The large dots indicate upper limits on $V_\phi$ from equilibrium constraints; the curves are continued to larger $V_\phi$ for purposes of illustration.

## 4.2.  *Critical Stability Curves*

In CPS, it was shown that $E$ behaves as

$$E = \frac{E_2(a)}{k^2} + \ldots + E_0(a), \tag{4.8}$$

where $E_0(a) \equiv r_1^{2a}$ and $E_2(a) \equiv \lim_{k \to 0} k^2 E$. Since the longest wavelength perturbations are always unstable in that case, one could then calculate the critical field strength for stability, $V_{z,crit}$, by taking the limit of the dispersion relation as $\omega \to 0$, $k \to 0$. When $V_\phi \neq 0$, the values of $(V_\phi, V_z)$ for which marginal stability holds constitute *curves* in the $(V_\phi, V_z)$ plane. This section will be concerned with the construction of such curves.

When $V_\phi \neq 0$, there is an added complication. Fig. 3 shows that the most persistent unstable mode is not always that with $k \to 0$. Rather, the last unstable mode which persists as $V_\phi \to 1$ has intermediate $k$; the precise value is a function of $V_z$. We note here that in the local limit, $k \to \infty$, $V_z \to 0$, the growth rate curves are perfectly symmetrical about $\Omega_A = 4$, which is the value for peak growth when $a = 2$ and $V_\phi = 0$.

When $V_\phi = 1$, equation (4.4) has the four roots

$$\Omega_A \text{ (twice)}, \quad \pm 2E^{-1/2} - \Omega_A, \tag{4.9}$$

all of which are real, provided that $E$ is real. This result is independent of both $V_z$ and $k$. Thus the line $V_\phi = 1$ must lie in an absolutely stable region in the $(V_\phi, V_z)$ plane. Further, taking $V_\phi = 1 \pm \epsilon$ with $\epsilon$ small and positive, and expanding $\omega$ in orders of $\epsilon$, one finds from the first-order correction that as long as $a < 2$, instability occurs. Thus *the line $V_\phi = 1$ constitutes an absolutely stable region in the $(V_\phi, V_z)$ plane.* We need not take any special care when considering $V_z \to 0$.

For larger values of $V_z$, say $V_z \gtrsim 0.3$, the limit $k \to 0$ *does* give a reliable estimate of the critical curve (Fig. 3a). Now taking

$$\omega = k\omega_1 + k^2 \omega_2 + k^3 \omega_3 + \ldots$$

along with equation (4.8), equation (4.4) becomes, to first order,

$$E_2 \omega_1^4 - 2[E_2 V_z^2 + a(V_\phi^2 - 1) + 2]\omega_1^2 + 8V_z V_\phi \omega_1 + V_z^2[EV_z^2 + 2a(V_\phi^2 - 1) - 4V_\phi^2] = 0. \tag{4.10}$$

Solving equation (4.10) for the loci of $\omega_1 = 0$ in the $(V_\phi, V_z)$ plane gives the critical stability curves we seek. These have been plotted in Figure 6. The $V_\phi = 0$ results, which were derived in CPS, are obtained where curves intersect the $V_z$ axis. One sees that as $V_\phi$ is increased from zero, smaller values of $V_z$ are needed for stabilization, until at $V_\phi = 1$ complete stabilization occurs for all $a$. For $a = 2$, all $V_\phi \geq 1$ are completely stable; this is represented by the heavy line along the $a = 2$ curve and continuing along the $V_\phi$ axis from $V_\phi = 1$ to infinity. For $a < 2$, the plane above $V_\phi = 1$ is divided into an unstable part (adjoining the $V_\phi$ axis) and a stable part (to the right of a given critical curve). The unstable region at $V_\phi > 1$ extends to infinity and shrinks to zero size as $a \to 2$. Actually, the size of the unstable region for $a \neq 2$ depends on the particular value of $a$ (and when $a \neq b$, on $b$ as well). This is due to the equilibrium constraints placed on $V_\phi$ by Fig. 2b. The largest allowed $V_\phi$ for each $a$ has been indicated in Fig. 6 by a large dot on the appropriate critical curve. We extend the curves to higher values of $V_\phi$ merely to display their asymptotic behavior (see below); such large field values will not be attainable in reality.

### 4.3. *The Large-Field Instability*

The almost linear behavior of the curves in Fig. 6 at large $V_\phi$, $V_z$ is intriguing. In this limit, and again taking $k \to 0$, equation (4.4) gives

$$\omega_1^2 = \frac{\Omega_A^2 [E_2 \Omega_A^2 + 2(a-2)\Omega_\phi^2]}{2k^2 (E_2 \Omega_A^2 + a\Omega_\phi^2)},$$

where $\Omega_\phi \equiv k V_\phi$. This implies

$$V_z/V_\phi > (V_z/V_\phi)_{crit} \equiv \sqrt{2(2-a)/E_2} \tag{4.11}$$

for stability. Values of $E_2$ and $(V_z/V_\phi)_{crit}$ for $3/2 \leq a \leq 2$ may be found in Table 1. The equality in (4.11) gives the asymptotic (i.e. large $V_\phi$, $V_z$) behavior of the critical stability curves, as shown by the dotted lines in Fig. 6.

The nature of this large-field instability (LFI) is easily understood upon comparison with the equivalent nonrotating system. The equilibrium pressure distributions are compared in Appendix B, where it is shown that *when $V_\phi \gg r_0 \Omega_0$, the system reduces to its nonrotating equivalent*. For the latter, Chandrasekhar (1961) derived the necessary and sufficient stability criterion

$$I_1 B_z^2 > \int_{r_1}^{r_2} \frac{\xi_r^2}{r^2} \frac{d}{dr} (r B_\phi)^2 dr, \tag{4.12}$$

| $a$ | $E_2(a)$ | $(V_z/V_\phi)_{crit}$ |
|-----|----------|----------------------|
| 1.5 | 2.55 | 0.63 |
| 1.6 | 2.78 | 0.54 |
| 1.7 | 3.01 | 0.45 |
| 1.8 | 3.26 | 0.35 |
| 1.9 | 3.50 | 0.24 |
| 2.0 | 3.75 | 0. |

Table 1: Ratio of critical Alfvén speeds, $(V_z/V_\phi)_{crit}$, as a function of shear parameter $a$ for the LFI. See text for the definition of $E_2(a)$.

where $I_1$ is a positive-definite integral function of $r$.[3] Differentiating both sides of this inequality, and assuming that $k \gg \partial/\partial r$ (this is equivalent to considering the longest-wavelength radial perturbations, which should be the most unstable), we obtain its local version:

$$k^2 B_z^2 > \frac{1}{r^3} \frac{d}{dr}(rB_\phi)^2 = \frac{2B_\phi}{r} J_z, \qquad (4.13)$$

where $J_z$ is the axial current. The LHS of (4.13) represents the restoring force exerted on a radially displaced fluid element by the perturbed vertical field, while the RHS is the excess Lorentz force on that element due to perturbations of $B_\phi$. *The latter is the exact analogue of the destabilizing centrifugal force in the BH instability.* Since $J_z = (2 - b)B_\phi/r$, configurations with $V_\phi \gg r_0\Omega_0$ and $b \geq 2$ are stable to the LFI. In essence, the LFI is the result of an imbalance between radial gravity and a radially stratified, buoyant magnetic field (see also Appendix B).

### 4.4. *Free Boundaries*

The only case to be considered here is $a = b = 2$, since we restrict consideration to the current-free situation. The critical stability curve is shown in Figure 7. The most significant difference is *the disappearance of the absolutely stable line at $V_\phi = 1$.* A glance back at the roots (4.9) of the polynomial (4.4) shows how this happens. When $V_\phi \neq 0$, $E$ is no longer real, and one of these roots becomes growing unstable. The actual behavior is as follows. Consider a line of constant $V_z$, such that $0 < V_z < 1$. The peak growth rate for a given $V_\phi$ decreases from a maximum at $V_\phi = 0$, to some minimum in the vicinity of $V_\phi \approx 1$, and then increases again without bound as $V_\phi$ is made larger. Note also how much more extended is the unstable region in the free case versus the rigid one.

Global effects must clearly be at work here, since $V_\phi > 1$ is unstable only in the free-boundary case. As rotation is not likely to be important in this region, it is instructive to consider the equivalent nonrotating problem. A situation similar, although not identical, to the latter is that of the plasma "pinch" (e.g., Chandrasekhar 1961, Ch. XII, §115). This consists of a filled cylindrical column of plasma, threaded by a uniform $B_z$, and surrounded by a vacuum region containing the same $B_z$ together with an azimuthal field $B_\phi \propto r^{-1}$. The entire arrangement is usually encircled by a concentric conducting wall, but we are free to place this at infinity and so ignore it for our present purpose.

---

[3]Compare criterion (4.12) with that for the VC instability; i.e.

$$I_1 V_z^2 > -\int_{r_1}^{r_2} (r\xi_r)^2 \frac{d\Omega^2}{dr} dr.$$

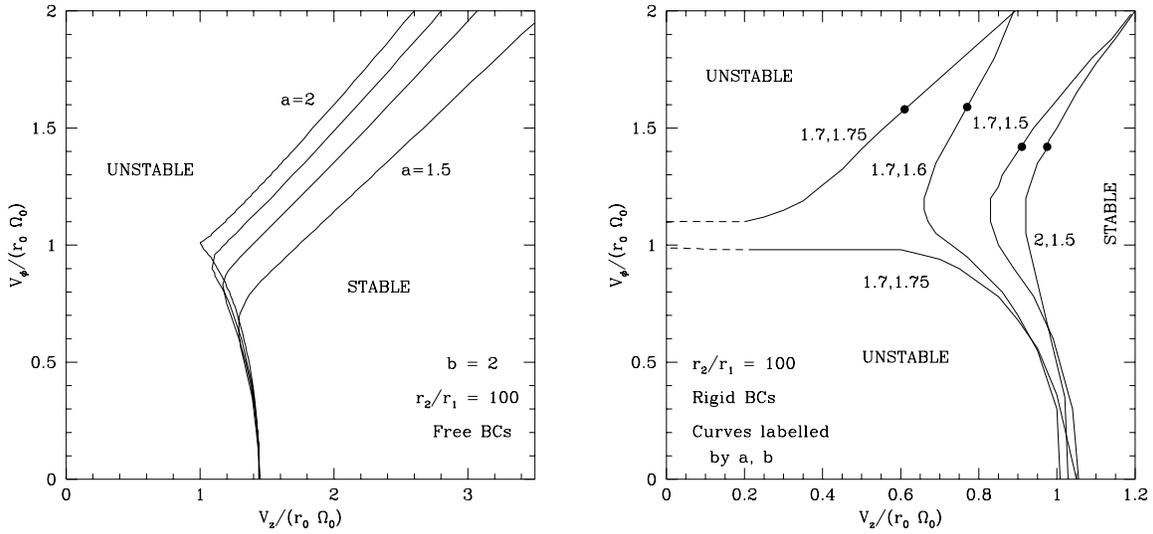

Fig. 7.— Same as Fig. 6, but for free BCs. From lower right to top left, curves are for $a = 1.5$, 1.7, 1.85, and 2. The region to the left of each curve is unstable; that to the right, stable.

Fig. 8.— Same as Fig. 6, but for $a \neq b$. Each curve is labelled by its corresponding $a$, $b$. For $a > b$, unstable regions lie to the left of each curve, stable regions to the right. For each $a < b$, there are two branches of the critical stability curve. One, at $V_\phi < 1$, bounds the VC unstable region from above; the other, at $V_\phi > 1$, bounds the large-field unstable region from below. The large dots indicate upper limits on $V_\phi$ from equilibrium constraints.

For the extended configurations we consider ($r_2/r_1 = 100$), the (nonrotating) situation is nearly identical except for the fact that in our problem $B_z$ and $B_\phi$ interpenetrate *everywhere*, not just in the vacuum region. However, such interpenetrating fields have been considered by Tayler (1957), with the finding that such arrangements are more unstable.

When all fields are continuous across the plasma/vacuum boundary, the fluid is susceptible to the well-known ($m = 0$) sausage instability, which can be stabilized if and only if $V_z^2 > V_\phi^2/2 \implies (V_z/V_\phi)_{crit} \gtrsim 0.707$. It is of interest to compare this figure with the inverse of the slope of the critical curve for $a = b = 2$ in Fig. 7, which is $(V_z/V_\phi)_{crit} \approx 1.5$. The latter situation is more unstable, we posit, due to the interpenetration of $B_\phi$ and $B_z$ in the fluid region. Since the exterior $B_\phi$ is the cause of the sausage instability in the first place, it is not hard to imagine that its presence *inside* the fluid will inhibit the stabilizing effect of $B_z$.

### 4.5. The General Case: $a \neq b$

### *4.5.1. Rigid Boundaries*

When $a \neq b$, the reduction of the full eigenvalue problem, equations (3.2) and (3.3), to a single characteristic polynomial is no longer possible. Before proceeding to a numerical solution, however, it is of use to present such analytic formulae as are available. There are two approaches which have had some success in this regard, and which lead to identical results. One is the local analysis of Dubrulle & Knobloch (1993), which ignores radial variations in equilibrium quantities compared with those of perturbed ones (i.e. $r(\delta X)'/\delta X \gg 1$). The other is the slender annulus approximation adopted by Kumar, Coleman, & Kley (1994) which we follow here to preserve the global character of the analysis (Appendix C).[4] In the limit $V_z \to 0$, both give the following condition for stability,

$$[2 - a - (2 - b)V_\phi^2](a - bV_\phi^2) < 0. \tag{4.14}$$

For example, if $0 < b < 2$ and $b < a$, stability holds if

$$\frac{2-a}{2-b} < V_\phi^2 < \frac{a}{b}, \tag{4.15}$$

whereas for $b$ in the same range and $a < b$, stability holds if

$$\frac{a}{b} < V_\phi^2 < \frac{2-a}{2-b}. \tag{4.16}$$

It is easy to see that both of these inequalities bracket $V_\phi = 1$.

As regards the $(V_\phi, V_z)$ critical stability plane, equations (4.15) and (4.16) imply the existence of a stable region along the $V_\phi$ axis bracketing $V_\phi = 1$. How this limiting behavior is related to the critical curves for general $V_\phi$, $V_z$, and $r_2/r_1$ will now be investigated.

For configurations with rigid boundaries, all $(a, b, V_\phi)$ consistent with Fig. 1b may be considered. Qualitatively, there are some significant differences from the $a = b$ case. These differences may be classified according as $a > b$ or $a < b$. Several representative critical stability curves are shown in Figure 8. Beginning at the far right-hand side of the diagram, we have a stable region at large $V_z$. When $a > b$, the curves achieve a minimum value of $V_z$ for some $V_\phi \gtrsim 1$, and then display the linear asymptotic behavior found in the previous section. *For $a > b$, there exists no stable region (not even $V_\phi = 1$) at small $V_z$.* This result contradicts the local prediction (4.15) of a stable region as $V_z \to 0$. As $a$ is reduced to values nearer to $b$, e.g. $a = 1.7$, $b = 1.6$, the "knee" of the curve bends inward to smaller values of $V_z$; it is easy to imagine what happens in the limit as $a \to b$ from above; the knee of the $a > b$ curve deforms into the *line $V_\phi = 1$*, which extends all the way to $V_z = 0$ as in Fig. 6.

---

[4] This approach is actually superficially global, in that although radial BCs are applied, the authors assume that the boundary separation is proportional to $\sqrt{k}$, $k \gg 1$.

If $a$ is decreased further such that $a < b$, the situation is less clear. We have been able to confirm numerically the persistence of two distinct unstable regions, one at $V_\phi \gtrsim 1$ (LF unstable) and one at $V_\phi \lesssim 1$ (VC unstable), down to values of $V_z \simeq 0.2$. *Between the two stability curves lies an absolutely stable region*, bracketing $V_\phi = 1$. At smaller $V_z$, mode crossing becomes a significant hindrance to the numerical algorithm, and precise determination of the critical curves is difficult. For $a = 1.7$, $b = 1.75$, we were able to follow the $n = 0$ mode down to $V_z \simeq 0.2$ (solid curves in Fig. 8); beyond this, we join the numerical curves onto the values given by the local relation (4.16) at $V_z = 0$ (dashed curves).

It should be mentioned that this region of parameter space, i.e.,

$$V_z \to 0, \ V_\phi \approx 1, \ a < b,$$

is highly restricted by the equilibrium constraints. A glance at Fig. 1a reveals that we must have $a \geq 3/2$. Since the LFI requires $b < 2$, we therefore have $3/2 < a < 2$, $a < b$ as our region of interest. Widely separated values of $a$ and $b$ in this range have limiting Alfvén speeds well below unity; e.g. when $a = 1.55$, $b = 1.95$, $V_6 = 0.46$. Hence, the LFI is not a concern. Less separated values of the two parameters allow larger equilibrium fields; e.g., $a = 1.85$, $b = 1.95 \Rightarrow V_6 = 2.93$. But it is likely that for such $a$, $b$ the critical stability curves are qualitatively similar to the $a = 1.7$, $b = 1.75$ case shown in Fig. 8. To confirm this, we developed an approximation whose validity depends on the smallness of the parameter $a/b$, but imposes no restrictions whatsoever on the global geometry.[5] The critical stability curves found by this method *always contain a stable region bracketing* $V_\phi = 1$.

To explain the existence of a stable region at small $V_z$, it is instructive to look at the dependence of the perturbations on $a$ and $b$. The VC instability arises from an imbalance of the destabilizing stress $B_z \delta B_\phi / 4\pi$ and the stabilizing stress $B_z \delta B_r / 4\pi$. When an azimuthal field is present, the ratio of these as found from equations (A.7) and (A.8) is

$$\frac{\delta B_\phi}{\delta B_r} = \frac{i\omega}{k B_z \tilde\omega^2} \left[ 2k B_z \Omega - \frac{B_\phi}{r\omega}(b\tilde\omega^2 + 2\Omega_A^2) \right]. \tag{4.17}$$

The first term in the square brackets behaves as $r^{-a}$, the second term as $r^{-b}$. Consider unstable modes only, so that $\omega \sim \Omega_A$ (this is still true when $V_\phi \lesssim 1$). The relative magnitude of the two terms then depends on: (a) the relative magnitude of $a$ and $b$, (b) the relative magnitude of $\Omega$ and $B_\phi/r$, and (c) whether $r < 1$ or $r > 1$ (i.e. inside or outside the pressure maximum, respectively). Assume that $\Omega \gtrsim B_\phi/r$; i.e. that we are in the VC regime. Recall from CPS how strongly peaked were the radial eigenfunctions of the unstable modes interior to the pressure maximum; this suggests that the region $r < 1$ is far more important than $r > 1$ for the linear stage of instability. We therefore restrict consideration to that region. Now, when $a > b$, the first term in the above dominates the second, and $\delta B_\phi / \delta B_r$ retains the *same* sign as it had in the absence of $B_\phi$, where its effects were always destabilizing. Thus while one would expect a reduction in the growth rate near $V_\phi \simeq 1$, it should not completely vanish.

---

[5]Specifically, we define a new variable, $x \equiv 1 - r^{a/b-1}$, $a < b$, and expand the perturbation equation (3.4) in powers of $x$. Finding a series solution and subjecting it to rigid BCs, one obtains a fourth-order dispersion relation similar to equation (4.10), which can be solved numerically for $\omega$. See Curry (1995) for details.

On the other hand, when $a < b$, the second term in equation (4.17) can be comparable to the first even when $B_\phi / r \lesssim \Omega$, and so a change of sign in $\delta B_\phi / \delta B_r$ occurs at some $V_\phi \lesssim 1$, signifying stabilization. Such stabilization cannot be maintained at higher values of $V_\phi$, however, once the LFI begins to set in. This gives the upper boundary of the stable region. Physically, $B_\phi$ overwhelms $\Omega$ in the inner disk when $b > a$, leading to momentary stabilization until the field becomes so strong that rotation is no longer a viable means of support. At this point, the LFI takes over. Because the local analysis gives no information about the radial dependence of the eigenfunctions, Dubrulle & Knobloch were not able to detect this interesting dependence of the stability properties on the relative magnitudes of $a$ and $b$.

### 4.5.2. Inapplicability of the Local Approximation

There are two particular cases in which the local criterion predicts qualitatively different behavior than that examined above. When either $a = 2$ or $b = 2$, criterion (4.14) yields the following results:

(i) $a = 2$, $b > 2$; $V_\phi > \sqrt{2/b}$,
(ii) $a = 2$, $b < 2$; $V_\phi < \sqrt{2/b}$,
(iii) $b = 2$, $a < 2$; $V_\phi > \sqrt{a/2}$,
(iv) $b = 2$, $a > 2$; $V_\phi < \sqrt{a/2}$.

In all these cases, there exists only a single critical curve. Since (ii) and (iv) have $a > b$, we expect the local prediction to be unreliable by extension of the results of the previous section; thus we do not expect a critical stability curve to extend all the way to $V_z = 0$. In cases (i) and (iii), however, there is no a priori reason to doubt the local results.

As test cases, consider the physically interesting power law indices

(i) $a = 2$, $b = 3$; $\Rightarrow V_\phi > 0.82$,
(iii) $b = 2$, $a = 1.5$; $\Rightarrow V_\phi > 0.87$.

The first case is that of constant angular momentum, with a rapidly decreasing azimuthal field. The second is a zero-current, Keplerian configuration. By the results of

§4.3., both systems should be stable to the LFI. In addition, both the Michael (equation (1.1)) and the Howard & Gupta (equation (1.2)) criteria are satisfied. The equilibrium constraints place no restrictions on the value of $V_\phi$ for these $(a, b)$. The critical stability curves, calculated numerically, are shown in Figure 9. Again, it is difficult to extend the curves much past $V_z \lesssim 0.2$, but in case (i) we have been able (quite remarkably) to follow the curve down to $V_z = 0.05$. As in the rest of the paper, the results are for $n = 0$, which we have always found to be the fastest growing radial mode.

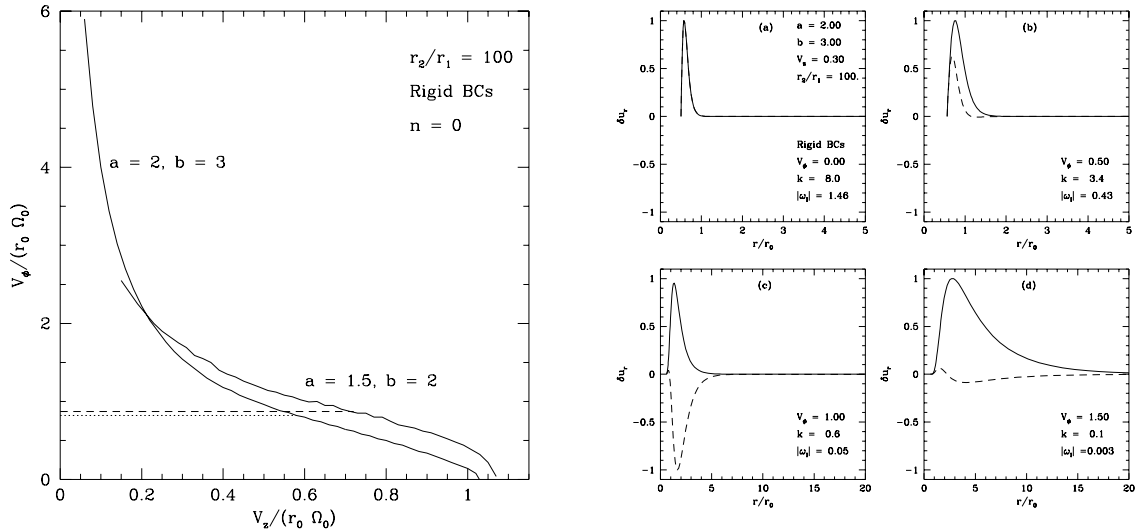

Fig. 9.— Critical stability curves for two special cases examined in §4.5.2. The dashed and dotted lines indicate the local predictions; they intersect the $V_\phi$ axis at $V_\phi = 0.87$ and $V_\phi = 0.82$, respectively.

Fig. 10.— Selected eigenfunctions at peak growth for $a = 2$, $b = 3$, and increasing $V_\phi$ from top left to lower right. The solid line indicates the real part of $\delta u_r$; the dashed line, the imaginary part. Each eigenfunction is normalized to its peak value.

The results are surprising in that they bear no resemblance to the local predictions (the dashed and dotted lines in Fig. 9) as $V_z \to 0$. The $a = 2$, $b = 3$ curve, e.g., shows that the VC unstable region is five times as large at $V_z = 0.1$ than the local prediction, and the curve even appears to be diverging as $V_z \to 0$, instead of approaching a constant value. One reason why the local approximation fails here can be found via inspection of the relevant eigenfunctions, a few of which are plotted in Figure 10. When $V_\phi \neq 0$, $\delta u_r$ has both real (solid line) and complex (dashed line) components. As $V_\phi$ is increased from zero, one sees a gradual spreading of the eigenfunction from the inside regions outward. In the region near the critical curves as $V_z \to 0$, $\delta u_r$ is *much* more extended than in any other case examined thus far. The peak of the eigenfunction at maximum growth is no longer confined to the small region between $r_1$ and $r_0$; e.g. when $V_z = 0.3$, $V_\phi = 1.5$ (Fig. 10d), it lies at $r/r_0 \approx 3$, and $\delta u_r$ has a nonnegligible amplitude over the entire shell. This feature alone is enough to show that the local and thin shell analyses are inadequate to capture the true behavior of the system in this parameter regime. It also confirms one of the main findings of CPS; namely, that the local and 'critical' limits are antipodal: the latter can only be reached via a global analysis.

### 4.5.3. Free Boundaries

For $b = 2$, we plot a variety of $a$ values in Fig. 7. Again, in contrast to the rigid BC case, *there exists no stable region around* $V_\phi = 1$; this is easily understood in light of the discussion given in §4.4. The unstable regions are larger for $a \neq 2$ than for $a = 2$: this is due to the fact that *two* instabilities, the current-driven LFI, and the sausage instability, act simultaneously. The asymptotic critical values for these curves range from $(V_z/V_\phi)_{crit} = 1.5$ for $a = 2$ to $(V_z/V_\phi)_{crit} = 2.3$ for $a = 1.5$. The $V_z$-axis intercepts of the curves match the values found in CPS.

### 4.6. The Effect of Simulated Vertical Boundaries

In CPS, we calculated the critical $V_z$ for several *fixed*, nonzero values of $k$, corresponding to vertical wavelengths, $\lambda_{crit}$, between 100 and 0.1 in units of the inner radius $r_1$. The intent was to gauge the probable effect of vertical disk boundaries on $V_{z,crit}$, under the hypothesis that the longest unstable wavelength could not exceed the disk thickness. The interesting result was that for $\lambda_{crit} = 0.1$, a reasonable value for a thin Keplerian disk, $V_{z,crit} \simeq 0.04 \approx V_{z,K}$, where $V_{z,K} \equiv \sqrt{6} \, c_s/\pi$ is the local Keplerian critical field estimate (BH). Thus, the super-rotational Alfvén speed required for stability in the infinite incompressible cylindrical shell model translates to a *super-thermal* $V_z$ in a thin, isothermal disk.

In the presence of an azimuthal field, we have found that for small $V_z$, values of $V_\phi \sim r_0 \Omega_0$ are required for critical stability. This therefore begs the same question as asked in CPS: does the same result hold for thin disks, or does critical stability again require $V_\phi \sim c_s$ ?

Following the same calculational procedure as in CPS, we calculated critical stability curves for $a = b = 2$, rigid boundaries, and a range of $\lambda_{crit}$ (Figure 11). Mode confusion prevents us from going to $\lambda_{crit} < 0.2$, but the trend is clear. The curves do not all approach $V_\phi = 1$ as $V_z \rightarrow 0$, since they are for *fixed* $k$; the small $\Omega_A$ stabilization discussed in §4.1. takes over when $V_z$ becomes small. This can be seen explicitly by deriving the following "local" critical stability relation. In the local limit, $E \rightarrow r_1^4$, so

$E^{1/2} \approx (0.5)^2 = 0.25$ for $a = 2$, and equation (4.6) gives (in proper units)

$$k_{crit} = \frac{2\pi}{\lambda_{crit}} = \frac{4\Omega_0}{V_z} \left\{ 1 \pm \left[ 1 - \left( \frac{V_\phi}{r_0 \Omega_0} \right)^2 \right]^{1/2} \right\}.$$

Assuming the azimuthal field is subthermal so that it does not significantly alter the overall structure of the disk, the critical stability requirement $\lambda_{crit} \approx 2H = 2\sqrt{2} c_s / \Omega(r_1)$ then yields

$$\left( \frac{V_\phi}{r_0 \Omega_0} \right)^2 \approx 1 - \frac{\pi^2}{2} \left( \frac{V_z}{c_s} - \frac{\sqrt{2}}{\pi} \right)^2 = 1 - \left( \frac{V_z}{r_0 \Omega_0} \frac{2\pi \, r_1}{\lambda_{crit}} - 1 \right)^2. \tag{4.18}$$

For $\lambda_{crit} = 0.1 \ r_1$, equation (4.18) gives the long-dashed curve shown in Figure 11. Although equation (4.18) concurs with the sequence of curves shown and highlights their key qualitative features, it cannot be rigorously correct for two reasons: first, one cannot actually have a "thin" disk with $a = 2$; and second, the derivation is inconsistent for $V_\phi / r_0 \Omega_0 \sim 1 \gg V_\phi / c_s$.

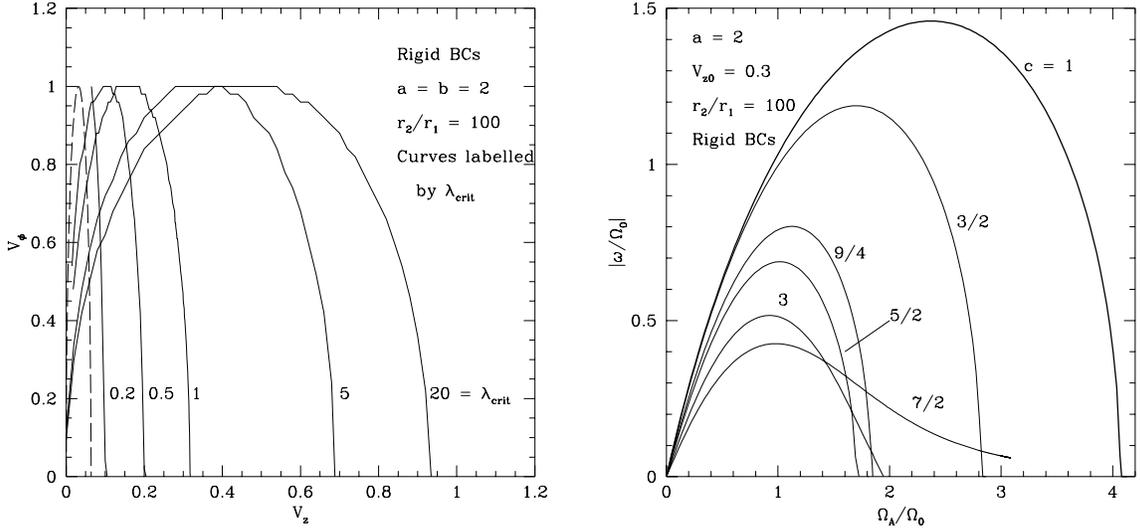

Fig. 11.— Critical stability curves for selected perturbation wavelengths $\lambda_{crit}$, for $a = b = 2$ and rigid BCs. The long-dashed curve is the "local" critical curve given by equation (4.18) with $\lambda_{crit} = 0.1 \ r_1$.

Fig. 12.— Numerical growth rates as a function of Alfvén frequency for $V_\phi = 0$ and $V_z = V_{z0} r^{1-c} (n = 0$ mode). Curves are labelled by their corresponding $c$ values.

Irregardless of the applicability of equation (4.18), the numerical curves in Figure 11 unambiguously show that although $V_{z,crit}$ decreases with decreasing $\lambda_{crit}$ (or decreasing scale height $H$), *the same is not true of* $V_{\phi,crit}$. Even in the thin disk limit, one still requires $V_{\phi,crit} \sim r_0 \Omega_0$ for complete stabilization; i.e. for all wavelengths and at any $V_z$. This result can be understood by recalling the physical cause of the LFI: it can *only* occur when rotation is relatively unimportant in comparison with the azimuthal field, a requirement that does not change when the effective scale height is reduced.

In a real, compressible, vertically stratified accretion disk, Parker (vertical magnetic buoyancy) instability is known to act when $V_\phi \gtrsim c_s \ll r_0 \Omega_0$. Thus, the above result could have at least two important consequences for such a disk. First, it argues persuasively against the possibility of the LFI ever occurring, since for $V_\phi \gtrsim c_s$, Parker instability would already have caused a rearrangement of the magnetic equilibrium. Second, and more importantly, the above result suggests that *the VC instability is unlikely to be stabilized by an azimuthal field of* any *power-law index or strength* $V_\phi \lesssim c_s$. We will discuss other possible environments for the LFI in §6.2.

## 5. NONCONSTANT VERTICAL FIELD

Should an accretion disk be threaded by a vertical magnetic field, the latter is more likely to vary with radius than be uniform. Although we do not explicitly model the accretion flow in this study, its overall effect is to drag field lines radially inward (by flux-freezing), leading to a higher $B_z$ flux in the inner regions. In this section we consider the effect of a radially varying vertical magnetic field on the VC instability, and neglect the azimuthal field. Although for completeness it would be desirable to consider the most general situation of nonconstant vertical *and* azimuthal fields, we defer that to a future work. An additional complication arises in that case, since resonances can occur where the real part of $\omega^2 - k^2 V_z^2(r) = 0$. This is not a concern for the unstable modes considered in this section, since they always have $\omega^2 < 0$; a proof of this is given in Appendix D. We consider only rigid BCs, since the zero-current restriction on our freely-bounded equilibria requires $c = 1$.

Even with the restriction to rigid BCs and the knowledge that $\omega^2$ is real, analytic progress is difficult, since the $r$-dependence of $\tilde{\omega}^2$ means that the perturbation equation (3.2) is not of standard Sturm-Liouville type. Regrettably, the global WKB approach used in CPS does not give satisfactory results in this case, for the following reason. Choosing $1/k$ as a small parameter, the last RHS term of equation (3.5) cannot be neglected, due to the presence of $\omega$-dependent terms. However, use of the thin shell approximation of Appendix C gives the result

$$
\begin{aligned}
\omega^2 \;=\; & \frac{\Omega_{A0}^2[2(k^2+c)+\eta/2]+2(2-a)k^2}{2(k^2+3/4)} \\
\pm\; & \{\Omega_{A0}^4[k^2(\eta+1)+\eta(\eta/4+2c-3/4)+3(1-2c)+4c^2] \\
+\; & 2k^2\Omega_{A0}^2[8k^2+3a+(2-a)(4c+\eta)]+4k^4(2-a)^2\}^{1/2}/2(k^2+3/4),
\end{aligned}
$$

where $\eta \equiv 3 - 8c + 4c^2$ and $\Omega_{A0} = kV_z(r_0)$. This solution can be regarded as quan-titatively valid only in a small neighbourhood of the pressure maximum. However, it exhibits roughly the same qualitative behavior as the exact numerical solutions discussed below. In addition, taking $k \gg 1$ leads to the local dispersion relation of CPS and BH.

Exact numerical growth rates as a function of $\Omega_{A0}$ for various values of $c > 1$ and the fiducial values $a = 2$, $V_{z0} = 0.3$, $r_2/r_1 = 100$ are plotted in Figure 12. The different curves are labelled by their corresponding $c$ values. *For $c > 1$, the growth rate is always reduced from its constant $V_z$ value.* The critical Alfvén frequency for stability, $\Omega_{A0,crit}$, decreases with increasing $c$, until at some critical value, $c \approx 2.5 - 3$, it begins to increase again. The peak growth rate, however, continues to decrease. We have difficulty finding $|\omega|$ for $c \gtrsim 3.5$ and large $\Omega_{A0}$, possibly due to the simultaneous presence of several unstable modes with the same growth rate.

The particular laws $c = 9/4$ and $c = 5/2$ correspond to the flux distributions for two popular centrifugally-driven wind models; Blandford & Payne (1982) and Pelletier & Pudritz (1992), respectively. As far as the *stability* of these distributions is concerned, there is no great distinction between either; both are VC unstable. One should note, however, that both models require $B_\phi \neq 0$; in the former $B_\phi \sim B_z$, while in the latter, $B_\phi \sim r^{-1}$. Thus while the results of the present paper suggest that $V_\phi \lesssim r_0\Omega_0$ will further stabilize, a calculation explicitly incorporating $B_\phi$ is still necessary.

The run of peak growth rate with $c$ is shown in Figure 13, for $V_z = 0.3$, $r_2/r_1 = 100$, and different values of $a$. The $a = 1.5$ curve is incomplete because $1 < c < 3/2$ is forbidden by the equilibrium (Fig. 2b). The large dot on the vertical axis shows the constant $V_z$ value (see Fig. 7a of CPS).

The physical reason for the stabilization observed here is the same as for an azimuthal field in the presence of a constant $V_z$, except that now the additional vertical motions are induced by the gradient of the vertical field (see equation (A.4)). As for the effect on the stability criterion, we advance the following argument. In the inner region of the disk, $V_z(r < r_0) > V_z(r_0)$. Thus, the *local* instability at $r < r_0$ will be attenuated compared to the constant $V_z = V_z(r_0)$ situation. By the same argument, the local growth rate should be enhanced outside the pressure maximum. However, the unstable eigenmodes are strongly peaked inside $r = r_0$, when $c \approx 1$; this region is more important for the action of the VC instability. Thus for $c \gtrsim 1$, the attenuation effect dominates, and the critical wavenumber for stability, $k_{crit} = \Omega_{A0,crit}/V_{z0}$, is reduced from its value in CPS. For $c$ significantly greater than 1, an interesting phenomenon occurs (Figure 14). The peak of the eigenfunction $\delta u_r$ gradually moves from inside the pressure maximum (for $c \approx 1$) to $r/r_0 \approx 1$ (for $c = 7/2$), and presumably beyond for more extreme field gradients. Thus it is likely that for larger $c$, the above argument no longer holds. That is, the enhancement effect of the VC instability at $r > r_0$ *does* contribute, leading to a reversal in the trend of $\Omega_{A0,crit}$.

We have searched for other unstable modes, e.g. at $V_z \gg 1$, with no success.

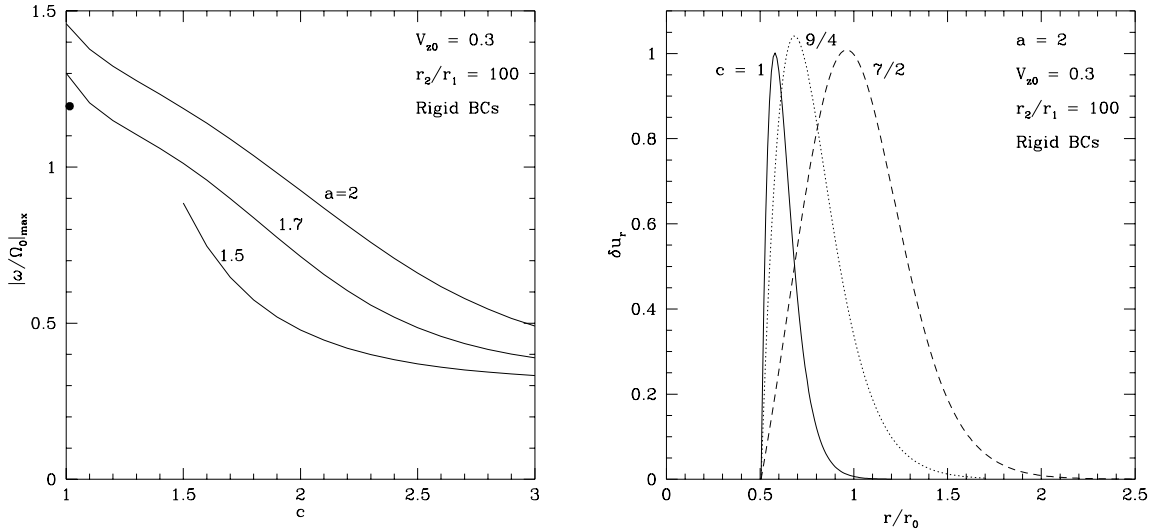

Fig. 13.— Maximum growth rates as a function of vertical field index $c$ for different rotation indices $a$. The lower curve is incomplete since equilibria with $a = 1.5$ and $1 < c < 1.5$ are not allowed (Fig. 2a). The maximum growth rate at $c = 1$, found in CPS, is shown as a large dot slightly offset (for clarity) from the vertical axis.

Fig. 14.— Radial eigenfunctions ($n = 0$ mode) at peak growth rate for various $c$. The overall normalizations have been adjusted to unity for purposes of comparison.

Interchange modes, which might be expected to act at large field strengths, do not occur here because we consider only axisymmetric perturbations (see, e.g., Kaisig, Tajima, & Lovelace 1992, Lubow & Spruit 1995).

# 6. DISCUSSION

## 6.1. *Comparison with Previous Results*

Here we compare our results for the effect of the azimuthal field on the VC instability with those of four recent papers, finding some significant discrepancies.

Dubrulle & Knobloch (1993) (DK), via a WKB method, found that the imaginary part of the eigenfrequency, $\omega_I \sim \Omega_A/(1 + \text{const.} \times V_\phi^2)$ in the limit $V_z \to 0$. The same result holds for both rigid and "free" BCs, $\delta u_r'(r_1) = \delta u_r'(r_2) = 0$ (these conditions differ from ours in the respect that the configuration is bounded by a *complete* vacuum; i.e. one devoid of external fields.) Thus it would appear that one needs an infinite $V_\phi$ to stabilize the system. Our results are clearly at odds with DK in this respect. Although the finite-sized stable region found by DK was also found here, we have shown that such a region exists *only* in the presence of rigid boundaries, and then only for $a < b$.

Kumar, Coleman, & Kley (1994) (KCK) concluded, on the basis of the sufficient stability criterion (1.2), that "toroidal fields only destabilize the flow". As regards the VC instability, we have found that the opposite is in fact the case, at least when we consider the "principal range" $3/2 \leq a \leq 2$, $b \geq 1$. It is only in the large-field ($V_\phi \gtrsim r_0\Omega_0$) regime that $B_\phi$ destabilizes. Had the authors continued their thin-shell calculation to $O(V_\phi^2)$, they would have discovered that the correction to $\omega$ at peak growth is

$$\omega_{2,max} = \frac{i}{2}(b - a^2/2 + a^3/8),$$

which is always damping provided that $3/2 \leq a \leq 2$ and $b \geq 1$.

As regards the enhancement of the instability for free boundaries, we note that the global energy change due to the perturbations, $\delta\mathcal{E}$, consists of three different contributions, in general. The first is the energy change in the fluid *interior*, derived by KCK as

$$\delta\mathcal{E}_F = \pi \int \left( \frac{|\delta\mathbf{B}|^2}{4\pi} - \mathbf{J}\cdot\delta\mathbf{B} \times \xi^* + 2\rho r\Omega\Omega'|\xi_r|^2 \right) r\,dr, \tag{6.1}$$

where $\xi = \delta\mathbf{u}/i\omega + r\Omega'\delta u_r\hat{\phi}$ is the Lagrangian displacement vector. The second contribution is due to perturbations of the external vacuum field,

$$\delta\mathcal{E}_V = \frac{1}{4} \int_{vacuum} |\delta\mathbf{B}|^2 r\,dr, \tag{6.2}$$

while the third is a surface contribution, $\delta\mathcal{E}_S$, which vanishes unless the equilibrium has surface currents (cf., Schmidt 1966). We avoid the latter here, and so the effect of free boundaries is given entirely by the integral (6.2), which is always positive. This led KCK to conclude that "stability criteria are not affected" by the BCs. However, one should be careful upon drawing such a conclusion from *sufficient*, but not *necessary*, criteria. In fact, as noted by Bateman (1978), there are numerous instances when free boundary instabilities grow faster than fixed boundary ones, even though $\delta\mathcal{E}_V > 0$. The reason for this is simply that by allowing $\xi \neq 0$ at the edge of the fluid, free-boundary instabilities can make more effective use of the *internal* fluid potential energy, represented by the first two terms on the RHS of equation (6.1). We found ample evidence of such behavior in the preceding sections, and in CPS.

Blaes & Balbus (1994) (BB) considered two-fluid models of ions and neutrals coupled by collisions, ionization, and recombinations. Their analysis is local, but includes an equilibrium azimuthal field. They found that $B_\phi$ can alter the stability criterion only in the limit of ionization equilibrium (as opposed to ion conservation), and can in fact produce total stabilization for $B_\phi \gtrsim 10B_z$ if the ion-neutral collision frequency is below a certain threshold. In all other cases, $B_\phi$ can cause a small reduction in growth rate, but does not affect the stability criterion (i.e. the critical Alfvén frequency for stability is unchanged from the $B_\phi = 0$ case)[6]. They take $c_s = 10V_z$, so that the critical $V_\phi$ for stability is $V_\phi \sim c_s$. This differs from our result, $V_{\phi,crit} \sim r_0 \Omega_0$, since BB's compressible model is sensitive to the coupling between magnetosonic and rotation-modified Alfvén modes, which is stabilizing. BB's model does not include vertical gravity, however, so buoyancy instabilities which would be expected to become important near $V_\phi \sim c_s$ were not detected.

---

[6]A point of formalism is worth stressing here. The finding that $B_\phi$ does not affect the stability criterion, regardless

Gammie & Balbus (1994) (GB) considered an accretion disk model which was local in the radial coordinate, but global in $z$; i.e. they solved for the vertical eigenmodes. One should be cautious in comparing our results directly to theirs, but their vertical node number $n$ should compare roughly with our $k$, and their radial wavenumber $k$ with our radial node number $n$. The near-coincidence of notation here is unfortunate; let us unambiguously re-label these parameters as $n_z$, $k_z$, $k_r$, and $n_r$, respectively. For a Keplerian disk, they plotted curves of constant growth rate in the $(V_\phi, V_z)$ plane for $k_r = 0$ and $n_z = 1$ (their Fig. 2), finding that stabilization is achieved for $V_z \simeq 1.5$ irrespective of $V_\phi$; this value agrees quite well with the free-boundary results of CPS (we found $V_{z,crit} \simeq 1.43$ for $a = 1.5$). Their BCs are similar to ours in the sense that far from the disk, the field lines move about freely, exerting no stress on the disk.

On the other hand, although GB find that the growth rate decreases for increasing $V_\phi$ (they consider values up to $V_\phi/c_s = 5$), it apparently never vanishes, nor does $V_\phi$ affect the stability criterion. The discrepancy between these results and those of the present paper could be telling us something about the relative importance of vertical motions (which they treat in detail, and we do not) and radial ones (vice-versa). To date, nonlinear calculations of the BH instability have indicated that inward and outward radial motions at different $z$ (the so-called "channel solutions") are the immediate outcome of the linear stage of the instability. It may be that the unstable modes are more sensitive to variations in radial structure than in vertical. GB's local approximation in $r$ could therefore have missed the most important effect associated with strong $B_\phi$; namely, the prevention of the channel solution from ever forming.

Due to the apparent similarity between GB's Fig. 2 and our Fig. 6, one might be tempted to make a direct comparison between the two. We caution the reader against it, for the following reason. The results in the former figure are for the longest vertical wavelength ($n_z = 1$ or $k_z = 0$) mode only. For this mode, the $V_z \to 0$ limit is automatically stable, since $\omega_I \sim \Omega_A \to 0$. By contrast, our critical stability curves are *mode-independent*; i.e. they reflect the requirements for stability to perturbations of *arbitrary* $k$. This explains the rather puzzling feature of GB's Fig. 2 in the $V_\phi \to 0$ limit, namely, that the absolute maximum growth rate is attained not at $V_z = 0$ as in CPS, but at $V_z \simeq 0.85$. As an example, consider the $a = b = 2$ case, whose growth rate is given by the imaginary part of equation (4.5). In the $k \to 0$ limit, $E \to E_2(2)/k^2$,

---

of its strength, is not surprising in a purely local model such as that of BB. This is because terms behaving as $B_\phi/r$, which are crucial in the global model, are ignored in local calculations. The disappearance of $B_\phi$ from the stability criterion in the latter case can be seen immediately from equations (4.7) and (4.17).

giving

$$\omega_I = -k(V_\phi^2 + E_2 V_z^2 - 2E_2^{1/2} V_z)^{1/2}/E_2.$$

Considered as a function of $V_z$, the maximum of $\omega_I$ occurs at $V_z = 0.52$, independent of $V_\phi$. The point this argument overlooks is that as $V_z$ becomes small, $k$ necessarily becomes large for the most unstable mode; e.g., when $V_z = 0.05$ and $V_\phi \gtrsim 0.7$, there are *no unstable modes* whatsoever at $k = 0$ (Fig. 3b). GB's Fig. 4 in fact shows that $n_z = 1$ is *not* the fastest growing mode for nonzero $V_\phi$. One should therefore not treat GB's Fig. 2 as our Fig. 6; i.e. as a critical stability diagram.

Finally, we note that in the context of uniformly rotating magnetic stars, which are expected to have distributions of $B_\phi$ *increasing* with radius, Pitts & Taylor (1985) identified an instability having the same characteristics as the LFI (i.e. stability was ensured for low $m$ (azimuthal wavenumber) modes provided that $r_0\Omega_0 \gtrsim V_{\phi 0}$), but did not obtain detailed growth rates or critical stability curves.

## 6.2.   *The Large-Field Instability: Possible Environments*

The results of §4.6. suggest that the LFI is not likely to be a threat in standard thin accretion disks. In some environments, however, the characteristic value for the LFI, $V_\phi/r_0\Omega_0 \gtrsim 1$, might in fact be achieved. Recent observations of flattened structures in massive star-forming regions (e.g., Aitken et al. 1993) suggest that such 'pseudo-discs' are very massive ($\sim 10^3 M_\odot$) and also that the dominant magnetic field component is toroidal. Such massive objects are likely to be self-gravitating and sub-Keplerian, so that rotation may not be as important a mechanism of support as in thin disks. It remains to be seen how the LFI is affected by self-gravity.

On larger scales, roughly 50 % of giant molecular clouds and somewhat fewer individual dark clouds and cores (Goldsmith & Arquilla 1985) possess measured velocity gradients which have been interpreted as being due, at least in part, to large-scale rotation (Blitz 1993). As the magnetic fields in such objects are substantial (magnetic energy $\sim$ gravitational energy $\sim$ kinetic (nonthermal) energy; cf., Myers & Goodman 1988), the condition $V_\phi/r_0\Omega_0 \gtrsim 1$ is likely to be satisfied in at least some regions. Of course, the effects of compressibility and self-gravity are also likely to be important, so a new model is needed.

A concrete example displaying appropriate conditions for the LFI may already exist. The L1641 region of Orion A consists of several low-density filaments, whose major axes run in a roughly north-south direction. In addition to a north-south velocity gradient which extends across all of Orion A ($\sim 8$ km s$^{-1}$), L1641 also contains an east-west gradient, $\sim 2$ km s$^{-1}$, indicating that the overall velocity field of Orion A is *helical* in nature (Bally 1989). Further, the surrounding magnetic field displays the same symmetry (Heiles 1987). It is well-known that such a helical field is characteristic of superposed vertical and azimuthal fields. While figures for L1641 alone are hard to come by, the average east-west gradient in the Orion A cloud as a whole (40 pc × 2 pc) has been estimated at 0.135 km s$^{-1}$ pc$^{-1}$ (Kutner et al. 1977, Genzel & Stutzki 1989). If this is entirely due to rotation of a cylindrical region $\simeq 20$ pc in radius, then a crude estimate of the rotation velocity gives $V_c \simeq 2.7$ km s$^{-1}$. Comparing this with $V_A \simeq 1.8$ km s$^{-1}$, the density-averaged Alfvén speed for the region (Heiles et al. 1993), one obtains $V_A/V_c \simeq 0.67$. Given the likelihood that $V_A \approx V_\phi$ (due to the predominantly toroidal appearance of the field), and that $V_c$ is probably a smaller contribution to the overall shear, one sees that values of $V_\phi/(r_0\Omega_0) \gtrsim 1$ should not be out of reach in this environment, and perhaps several others.

### 6.3.   *Summary*

In this paper we have examined a variety of magnetic field distributions and orientations, with the principal intent of gauging their effect on the VC instability of magnetized accretion disks. The main results are: (1) An azimuthal field, varying as some inverse power of radius, has a stabilizing effect on the VC instability if its characteristic Alfvén speed, $V_{\phi 0}$, is less than the characteristic rotational speed, $r_0\Omega_0$. (2) If $V_{\phi 0} \gtrsim r_0\Omega_0$, the system is susceptible to the LFI, whose peak growth rate increases with $V_{\phi 0}$. This instability is more likely to affect thick, massive disks and molecular clouds than thin accretion disks. (3) Our calculations for finite critical wavenumbers suggest that complete stabilization of thin disks by an equilibrium $B_\phi$ is unlikely, since the required field ($V_\phi \sim r_0\Omega_0 \gg c_s$) is prone to Parker instability. (4) In contrast to CPS, taking account of the disk's free boundaries gives qualitatively different behavior. In particular, whereas absolute stability can be achieved for certain rigidly-bounded configurations, none of the freely-bounded equilibria we examined are similarly stable. (5) In the absence of an equilibrium azimuthal field, a disk with a radially-varying vertical field has a smaller VC growth rate than in the constant field case. However, the most unstable wavenumber for fields which decrease extraordinarily quickly with radius may be unaffected or even increased.

The advantages of adopting a global analysis to address questions of stability in the presence of strong magnetic fields are even more apparent in the present work than in CPS. In particular, our results show that differentially rotating gaseous bodies threaded by strong azimuthal, but weak vertical fields should be highly unstable for certain specific rotational and azimuthal field profiles (§4.5.2.), a result not definitively shown by any local or thin shell analysis. It is hoped that future work will focus on these particular profiles, in order to more fully examine the consequences of the ensuing instabilities.

We thank Peter Sutherland for reading an earlier version of the manuscript and Omer Blaes for several useful discussions. C.C. is grateful to McMaster University for financial support, while the research of R.E.P. is supported by the Natural Sciences and Engineering Research Council of Canada.

## APPENDICES

## A. THE PERTURBATION EQUATIONS

We begin with the equations of ideal MHD in cylindrical polar coordinates $(r, \phi, z)$:

$$\rho \left[ \frac{\partial \mathbf{u}}{\partial t} + (\mathbf{u} \cdot \nabla) \mathbf{u} \right] = -\rho \nabla \Psi - \nabla \left( p + \frac{\mathbf{B} \cdot \mathbf{B}}{8\pi} \right) + \frac{1}{4\pi} (\mathbf{B} \cdot \nabla) \mathbf{B}, \quad (A.1)$$

$$\frac{\partial \mathbf{B}}{\partial t} = \nabla \times (\mathbf{u} \times \mathbf{B}), \quad (A.2)$$

$$\nabla \cdot \mathbf{u} = \nabla \cdot \mathbf{B} = 0. \quad (A.3)$$

Here $p$ is the gas pressure, $\rho$ the constant density, $\mathbf{u}$ the fluid velocity, and $\Psi = -GM/r$ the gravitational potential. Substituting perturbations of the form (3.1) into these equations and only retaining terms of linear order in perturbed quantities, we obtain

$$i\omega \delta \mathbf{u} + \nabla \delta h - \frac{ikB_z}{4\pi\rho} \delta \mathbf{B} + 2 \left( \frac{B_\phi}{4\pi\rho r} \delta B_\phi - \Omega \delta u_\phi \right) \hat{\mathbf{r}}$$

$$+ 2 \left( \frac{\overline{\mathcal{B}}}{4\pi\rho} \delta B_r - \mathcal{B} \delta u_r \right) \hat{\phi} - \frac{B_z'}{4\pi\rho} \delta B_r \hat{\mathbf{z}} = 0, \quad (A.4)$$

$$i\omega \delta \mathbf{B} - ikB_z \delta \mathbf{u} + 2(\mathcal{A} \delta B_r - \overline{\mathcal{A}} \delta u_r) \hat{\phi} + B_z' \delta u_r \hat{\mathbf{z}} = 0, \quad (A.5)$$

$$\frac{1}{r} (r \delta u_r)' + ik \delta u_z = 0, \quad (A.6)$$

where $h = p/\rho + B^2/(8\pi\rho)$ is the specific enthalpy, $\mathcal{A} = -r\Omega'/2$, $\mathcal{B} = -[(r\Omega)' + \Omega]/2]$ are the usual Oort shear parameters, and $\overline{\mathcal{A}} \equiv -r(B_\phi/r)'/2$, $\overline{\mathcal{B}} \equiv -(B_\phi' + B_\phi/r)/2$ are their magnetic counterparts. For future reference, we note that with the power-law forms (2.1), the first two of these equations become

$$i\omega \delta \mathbf{u} + \nabla \delta h - \frac{ikB_z}{4\pi\rho} \delta \mathbf{B} + 2 \left( \frac{B_\phi}{4\pi\rho r} \delta B_\phi - \Omega \delta u_\phi \right) \hat{\mathbf{r}}$$

$$+ \left[ (2-a)\Omega \delta u_r - \frac{(2-b)B_\phi}{4\pi\rho r} \delta B_r \right] \hat{\phi} - \frac{(1-c)B_z}{4\pi\rho r} \delta B_r \hat{\mathbf{z}} = 0, \quad (A.7)$$

$$i\omega \delta \mathbf{B} - ikB_z \delta \mathbf{u} + \left( a\Omega \delta B_r - \frac{bB_\phi}{r} \delta u_r \right) \hat{\phi} + (1-c)\frac{B_z}{r} \delta u_r \hat{\mathbf{z}} = 0. \quad (A.8)$$

Note that equations (A.5) and (A.6) imply $\nabla \cdot \delta \mathbf{B} = 0$. Resolving equations (A.7) and (A.8) into components, and using equation (A.6), one can eliminate all variables except $\delta u_r$, leading to the perturbation equation (3.2).

## B. ROTATING VS. NONROTATING EQUILIBRIA

The stationary pressure distribution in the nonrotating case can be found from equation (2.2) with $\Omega_0 = 0$. When $V_z$ = constant, the pressure maximum relation is simply

$$\frac{GM}{r_0} = (b-2)V_{\phi 0}^2. \tag{B.1}$$

Note that this requires $b > 2$ for a sensible equilibrium. Using this to eliminate $GM$ in equation (2.2) and integrating gives

$$\left.\frac{p}{\rho}\right|_{\Omega=0} = \frac{p_0}{\rho} + V_{\phi 0}^2(b-2)\left[\frac{r_0}{r} - 1 + \frac{1 - (r/r_0)^{-2(b-1)}}{2(b-1)}\right]. \tag{B.2}$$

For a constant vertical field, equation (2.5) reads (in proper units)

$$
\begin{aligned}
\frac{p}{\rho} \;=\; & \frac{p_0}{\rho} + (r_0\Omega_0)^2 \times \\
& \left\{\frac{r_0}{r} - 1 + \frac{1 - (r/r_0)^{-2(a-1)}}{2(a-1)} - (2-b)\frac{V_{\phi 0}^2}{r_0^2\Omega_0^2}\left[\frac{r_0}{r} - 1 + \frac{1 - (r/r_0)^{-2(b-1)}}{2(b-1)}\right]\right\}.
\end{aligned}
$$

When $b = a$, this becomes

$$
\begin{aligned}
\frac{p}{\rho} \;=\; & \frac{p_0}{\rho} + [(r_0\Omega_0)^2 - (2-b)V_{\phi 0}^2]\left[\frac{r_0}{r} - 1 + \frac{1 - (r/r_0)^{-2(b-1)}}{2(b-1)}\right] \\
\;=\; & \frac{p_0}{\rho} + V_{\phi 0}^2(b_{eff} - 2)\left[\frac{r_0}{r} - 1 + \frac{1 - (r/r_0)^{-2(b-1)}}{2(b-1)}\right], \tag{B.3}
\end{aligned}
$$

where

$$b_{eff} \equiv \left(\frac{r_0\Omega_0}{V_{\phi 0}}\right)^2 + b.$$

Clearly, equation (B.3) is identical to equation (B.2), but for the replacement of $b$ by $b_{eff}$. The two equations become identical in the limit $V_{\phi 0} \gg r_0\Omega_0$, which is precisely the regime of the LFI found in this paper. In fact, as soon as $V_{\phi 0} \gtrsim r_0\Omega_0$, one would expect that the rotating system should start to display much of the qualitative behavior of its nonrotating counterpart, since then the contribution of the magnetic terms to the pressure is of the same sign in equations (B.2) and (B.3) for $b \gtrsim 1$. Finally, one might be tempted to blame the LFI for $b < 2$ entirely on the violated equilibrium condition (B.1). However, this condition applies only when $V_{\phi 0} \gg r_0\Omega_0$. As an example, take $b = 1.7$. Then Fig. 1b shows that all equilibria with $0 \leq V_\phi \leq 1.83$ are allowed.

## C. THIN-SHELL APPROXIMATION

Following Kumar, Coleman, & Kley (1994), we adopt a thin shell approximation, in which the radial dependence of equilibrium quantities is ignored to first order, but their derivatives are not. The perturbation equation (4.1) then becomes

$$\frac{d^2\psi}{d\zeta^2} + Q_0\psi = 0, \tag{C.1}$$

where

$$Q_0 = \frac{2k^2 V_\phi^2}{\tilde{\omega}^4}(b\tilde{\omega}^2 - 2\Omega_A^2) + \frac{2k^2}{\tilde{\omega}^4}(2\omega^2 - a\tilde{\omega}^2) - \frac{8k^2\omega\Omega_A V_\phi}{\tilde{\omega}^4} - (k^2 + 3/4), \tag{C.2}$$

$\psi \equiv \sqrt{r}\,\delta u_r$, and $\zeta \equiv r - 1$. Since $Q_0$ is a constant, the solution of equation (C.1) is $\psi = c_1 \sin\sqrt{Q_0}\,\zeta + c_2 \cos\sqrt{Q_0}\,\zeta$; applying the rigid BCs then gives $Q_0 = (n\pi/s)^2$, where $s$ is the shell half-thickness and $n$ the radial mode number. Assuming $ks \gg n$, equation (C.2) yields the following characteristic polynomial:

$$\omega^4 + [2(a - 2 - bV_\phi^2) - 2\Omega_A^2]\omega^2 + 8\Omega_A V_\phi\omega + \Omega_A^2[\Omega_A^2 + 2(b-2)V_\phi^2 - 2a] = 0, \tag{C.3}$$

where the reader is reminded that all equilibrium quantities are to be evaluated at $r = r_0 = 1$.

The roots of equation (C.3), although calculable analytically, are algebraically complicated and do not give much physical insight. Kumar et al. (1994) adopted a procedure equivalent to expanding $\omega$ in powers of $V_\phi$, taking the latter as a small quantity. As our object is to obtain the critical stability curves, it is more useful for our needs to place no restriction on $V_\phi$; rather, we take $\Omega_A$ as a small parameter. Expanding $\omega$ as

$$\omega = \omega_0 + \Omega_A\omega_1 + \Omega_A^2\omega_2 + \cdots,$$

substituting into equation (C.3), and solving the resulting equation in orders of $\Omega_A$, we find there are two branches of the dispersion relation. One gives all real contributions to $\omega$; the other has $\omega_0^2 = 0$ and

$$\omega_1^2 = \frac{-2V_\phi \pm [b(b-2)V_\phi^4 + 2(a + b - ab)V_\phi^2 - a(2-a)]^{1/2}}{a - 2 - bV_\phi^2}. \tag{C.4}$$

Positivity of the square-root argument leads to the stability criterion (4.14).

## D.  PROOF THAT $\omega^2$ IS REAL WHEN $B_z = B_z(r)$ AND $B_\phi = 0$

The perturbation equation in this case is [equation (3.2)]:

$$\frac{1}{r}[r\tilde{\omega}^2(\delta u_r)']' + \left\{k^2 r\left[(\Omega^2)' - \frac{(V_z^2)'}{r}\right] + \frac{4k^2\omega^2\Omega^2}{\tilde{\omega}^2} - \tilde{\omega}^2\left(k^2 + \frac{1}{r^2}\right)\right\}\delta u_r = 0.$$

Multiplying through by $r\delta u_r^*$ (an asterisk denotes the complex conjugate) and integrating, one finds

$$\int_{r_1}^{r_2} \left\{ k^2 r \left[ (\Omega^2)' - \frac{(V_z^2)'}{r} \right] + \frac{4k^2\omega^2\Omega^2}{\tilde{\omega}^2} - \tilde{\omega}^2 \left( k^2 + \frac{1}{r^2} \right) \right\} r|\delta u_r|^2 dr + I = 0, \quad \text{(D.1)}$$

where

$$I \equiv \int_{r_1}^{r_2} \delta u_r^* [r\tilde{\omega}^2(\delta u_r)']' dr = r\tilde{\omega}^2(\delta u_r)'\delta u_r^* \big|_{r_1}^{r_2} - \int_{r_1}^{r_2} r\tilde{\omega}^2 |(\delta u_r)'|^2 dr. \quad \text{(D.2)}$$

The latter result is obtained via integration by parts.

We consider rigid BCs only, since we restrict consideration to $B_z = $ constant in the free boundary case to avoid currents (§ 2.2.1.). Applying $\delta u_r(r_1) = \delta u_r(r_2) = 0$ to equation (D.2), the first RHS term vanishes. Substituting the result back into equation (D.1) and taking the imaginary part of the entire expression gives

$$(\omega^2)_I \int_{r_1}^{r_2} \left\{ |(\delta u_r)'|^2 + \left[ \frac{4k^4}{|\tilde{\omega}^2|^2} V_z^2 \Omega^2 + k^2 + \frac{1}{r^2} \right] |\delta u_r|^2 \right\} r\,dr = 0,$$

where a subscript I indicates the imaginary part. The integrand is positive definite for all $r$, showing that $(\omega^2)_I = 0$.